\documentclass[twocolumn]{aastex63}

\newcommand{\todo}{\ifmmode \text{\color{purple}\Huge{\(\bullet\)}} \else {\color{purple}{\Huge$\bullet$}}\fi}
\newcommand{\finish}{\ifmmode \text{\color{blue}\Huge{\(\bullet\)}} \else {\color{blue}{\Huge$\bullet$}}\fi}

\newcommand{\Msun}{M_{\odot}}
\newcommand{\Mbh}{M_\mathrm{BH}}

\newcommand{\LEdd}{L_\mathrm{Edd}}

\newcommand{\LuvalphaOX}{L_\mathrm{2500\,\text{\AA}}}

\newcommand{\Lsoftx}{L_\mathrm{0.5-2\,keV}}

\newcommand{\Lhardx}{L_\mathrm{2-10\,keV}}
\newcommand{\NH}{N_\mathrm{H}}

\newcommand{\alphaox}{\alpha_\mathrm{OX}}

\newcommand{\msun}{M_{\odot}}
\newcommand{\mbh}{M_\mathrm{BH}}

\newcommand{\lbol}{L_\mathrm{AGN,bol}}

\newcommand{\lambdaedd}{\lambda_\mathrm{Edd}}

\newcommand{\av}{A_\mathrm{V}}
\newcommand{\R}{\mathcal{R_{\rm rest}}}

\newcommand{\rt}[1]{#1}

\usepackage{natbib}
\usepackage{amsmath}
\usepackage{multirow} 
\usepackage{ifthen} 
\usepackage{txfonts}

\received{\today}
\submitjournal{ApJ}

\shorttitle{X-ray and radio-loud super-Eddington quasar}
\shortauthors{Obuchi et al.}

\begin{document}



\title{Discovery of an X-ray Luminous Radio-Loud Quasar at $z=3.4$: A Possible Transitional Super-Eddington Phase}

\correspondingauthor{Sakiko Obuchi, Kohei Ichikawa}
\email{buchi-13526.cjl@fuji.waseda.jp, k.ichikawa@astr.tohoku.ac.jp}

\author{Sakiko Obuchi}
\affiliation{Department of Physics, 
Graduate School of Advanced Science and Engineering, Faculty of Science and Engineering, Waseda University, 3-4-1,
Okubo, Shinjuku, Tokyo 169-8555, Japan}

\author[0000-0002-4377-903X]{Kohei Ichikawa}
\affiliation{Department of Physics, 
Graduate School of Advanced Science and Engineering, Faculty of Science and Engineering, Waseda University, 3-4-1,
Okubo, Shinjuku, Tokyo 169-8555, Japan}
\affiliation{Frontier Research Institute for Interdisciplinary Sciences, Tohoku University, Sendai, Miyagi 980-8578, Japan}
\affiliation{Astronomical Institute, Tohoku University, Aramaki, Aoba-ku, Sendai, Miyagi 980-8578, Japan}

\author[0000-0002-9754-3081]{Satoshi Yamada}
\affiliation{Frontier Research Institute for Interdisciplinary Sciences, Tohoku University, Sendai, Miyagi 980-8578, Japan}
\affiliation{Astronomical Institute, Tohoku University, Aramaki, Aoba-ku, Sendai, Miyagi 980-8578, Japan}

\author[0000-0003-2535-5513]{Nozomu Kawakatu}
\affiliation{Facutly of Natural Sciences, National Institute of Technology, Kure College, 2-2-11 Agaminami, Kure, Hiroshima, 737-8506, Japan}

\author[0000-0002-2941-6734]{Teng Liu}
\affiliation{Department of Astronomy, University of Science and Technology of China, Hefei 230026, People's Republic of China}
\affiliation{School of Astronomy and Space Science, University of Science and Technology of China, Hefei 230026, People's Republic of China}

\author[0000-0002-8299-0006]{Naoki Matsumoto}
\affiliation{Astronomical Institute, Tohoku University, Aramaki, Aoba-ku, Sendai, Miyagi 980-8578, Japan}

\author[0000-0002-0761-0130]{Andrea Merloni}
\affiliation{Max-Planck-Institut für Extraterrestrische Physik (MPE), Gießenbachstraße 1, 85748 Garching bei München, Germany}

\author[0009-0009-8116-0316]{Kosuke Takahashi}
\affiliation{Astronomical Institute, Tohoku University, Aramaki, Aoba-ku, Sendai, Miyagi 980-8578, Japan}

\author[0000-0002-5208-1426]{Ingyin Zaw}
\affiliation{New York University Abu Dhabi, Science Division, Physics Program, P.O. Box 129188, Abu Dhabi, UAE}

\author[0000-0003-2682-473X]{Xiaoyang Chen}
\affiliation{Frontier Research Institute for Interdisciplinary Sciences, Tohoku University, Sendai, Miyagi 980-8578, Japan}

\author[0000-0001-6906-772X]{Kazuhiro Hada}
\affiliation{Graduate School of Science, Nagoya City University, Yamanohata 1, Mizuho-cho, Mizuho-ku, Nagoya 467-8501, Aichi, Japan}
\affiliation{Mizusawa VLBI Observatory, National Astronomical Observatory of Japan, 2-12 Hoshigaoka, Mizusawa, Oshu, Iwate 023-0861, Japan}

\author[0000-0001-9274-1145]{Zsofi Igo}
\affiliation{Max-Planck-Institut für Extraterrestrische Physik (MPE), Gießenbachstraße 1, 85748 Garching bei München, Germany}
\affiliation{Exzellenzcluster ORIGINS, Boltzmannstr. 2, 85748, Garching, Germany}

\author[0000-0002-2536-1633]{Hyewon Suh}
\affiliation{International Gemini Observatory/NSF NOIRLab, 670 N. A'ohoku Place, Hilo, HI, 96720, USA}

\author[0000-0003-0643-7935]{Julien Wolf}
\affiliation{Max-Planck-Institut für Extraterrestrische Physik (MPE), Gießenbachstraße 1, 85748 Garching bei München, Germany}
\affiliation{Exzellenzcluster ORIGINS, Boltzmannstr. 2, 85748, Garching, Germany}
\affiliation{Max-Planck-Institut für Astronomie, Königstuhl 17, 69117 Heidelberg, Germany}



\begin{abstract}
We report the multiwavelength properties of eFEDS J084222.9+001000 (hereafter ID830), 
a quasar at $z=3.4351$, identified as the most X-ray luminous radio-loud quasar in the eROSITA
Final Equatorial Depth Survey (eFEDS) field. 
ID830 shows a rest-frame 0.5--2~keV luminosity of $\log (\Lsoftx/\mathrm{erg}~\mathrm{s}^{-1}) = 46.20 \pm 0.12$,
with a steep X-ray photon index ($\Gamma =2.43 \pm 0.21$), and a significant radio counterpart detected with VLA/FIRST 1.4~GHz and VLASS 3~GHz bands.
The rest-frame UV to optical spectra from SDSS and Subaru/MOIRCS $J$-band show a dust reddened quasar feature with $\av = 0.39 \pm 0.08$~mag and the expected bolometric AGN luminosity from the dust-extinction-corrected UV luminosity reaches $L_\mathrm{bol,3000\,\text{\AA}}= (7.62 \pm 0.31) \times 10^{46}$~erg~s$^{-1}$. 
We estimate the black hole mass of $\Mbh = (4.40 \pm 0.72) \times 10^{8} \msun$ based on the MgII$\lambda$2800 emission line width, and an Eddington ratio from the dust-extinction-corrected UV continuum luminosity reaches $\lambda_\mathrm{Edd,UV}=1.44 \pm 0.24$ and $\lambda_{\mathrm{Edd,X}} = 12.8 \pm 3.9$ from the X-ray luminosity, both indicating the super-Eddington accretion.
ID830 shows a high ratio of UV-to-X-ray luminosities, $\alphaox=\rt{-1.20 \pm 0.07}$ (or $\alphaox=\rt{-1.42 \pm 0.07}$ after correcting for jet-linked X-ray excess), 
higher than quasars and little red dots in super-Eddington phase with similar UV luminosities, with $\alphaox<-1.8$.
Such a high $\alphaox$ suggests the coexistence of a prominent radio jet and X-ray corona, in this high Eddington accretion phase.
We propose that ID830 may be in a transitional phase after an accretion burst, evolving from a super-Eddington to a sub-Eddington state, which could naturally describe the high $\alphaox$.
\end{abstract}




\section{Introduction}\label{sec:intro}

How supermassive black holes (SMBHs) acquired their mass, along with the stellar mass assembly of their host galaxies, 
remains one of the fundamental open questions in modern astronomy \citep[e.g.,][]{hic09,mad14,ued14,hec14,hec23,som15}.
The discovery of quasars at $z \approx 7.6$ with SMBH masses of $\mbh \approx 10^9 \msun$ \citep{banados2018,Yang2020,wang2021} already challenges standard black hole (BH) seed formation scenarios. 
In the case of population~III remnants \citep[e.g.,][]{mad01}, which are expected to produce BH seeds with masses of $M_\mathrm{BH}=10$--$1000$~$\msun$ \citep[e.g.,][]{hir14,hir15}, such massive SMBHs require 
continuous Eddington limit accretion for at least 650 Myr in a gas-rich environment, assuming a radiative efficiency of $\sim 10\%$ \citep[e.g.,][]{banados2018,ina20}.
Recent JWST discoveries have pushed the redshift of active galactic nuclei (AGN) up to $z\approx11$, further deepening the tension between observations and theoretical models of BH seed formation \citep[e.g.,][]{lar23,gou23,mai24,bog24,tay25}.

One possible solution is the occurrence of accretion phase exceeding the
classical Eddington limit. 
Radiative hydrodynamical simulations suggest that anisotropic, super-Eddington accretion can occur under certain conditions \citep[e.g.,][]{abr88,ohs05,ohs09,mck14,ina16,tak20},
and \rt{observational studies have reported candidates undergoing such super-Eddington phase \citep[e.g.,][]{Luo2015,du15,Harikane2023,Maiolino2024,suh25}. In paticular, \cite{suh25} reported a JWST-detected super-Eddington SMBH at $z\approx4$ with bright X-ray emission, whereas high or super-Eddington accretion typically exhibits X-ray weakness, both theoretically and observationally \citep[e.g.,][]{Luo2015,Zappacosta2020,Inayoshi2024,Pacucci2024}. 
\cite{Ighina2025} also reported an X-ray bright radio-powerful quasar at $z=6.13$, which is one of the most luminous quasars at $z>5.5$, and suggested that its remarkably soft and strong X-ray emission indicates super-Eddington accretion.}
While super-Eddington accretion may be preferentially triggered in the early universe, these episodes likely occur in environment surrounded by copious amounts of gas and dust \citep[e.g.,][]{Vito2019,eil21}. 
Such environment can heavily obscure the central SMBHs, resulting in redder quasar colors and making them easily missed in optical surveys that assume blue, unobscured quasar templates \citep{Kato2020,mat25}. 

Under this situation, multiwavelength AGN surveys become essential. 
By utilizing energy bands less affected by the effect of dust reddening and/or gas absorption, 
these surveys can identify AGN that are missed in the optical. 
X-ray and radio band surveys are particularly powerful in this aspect, 
since both bands offer strong penetrating capabilities, with X-rays (2--10~keV) able to probe gas absorption up to  $\NH<10^{24}$~cm$^{-2}$ \citep{Ueda2014,ric17} and GHz-frequency radio bands remaining detectable even through $\NH<10^{26}$~cm$^{-2}$ \citep[e.g.,][]{pad17,gil22,maz24,fuk25}.

Recently, 
\cite{Ichikawa2023} constructed the radio AGN catalog in the eROSITA Final Equatorial Depth Survey (eFEDS) field \citep{Predehl2021,Brunner2022,mer24}, through cross-matching between the VLA/FIRST \citep{Becker1994,hel15} and eROSITA X-ray source catalogs \citep{liu22,sal22}. 
Thanks to the wide survey area of eFEDS ($\sim140$~deg$^2$),
\cite{Ichikawa2023} identified several rare classes of AGN, including radio-loud obscured quasars and X-ray and radio-luminous AGN exceeding the knee of the luminosity functions at $1<z<4$.

Among these sources, one particularly interesting radio-loud AGN 
is eFEDS~J084222.9+0010000 (cataloged as eFEDS object ID=830, we hereafter call the target ID830) at $z=3.4351$.
Figure~\ref{fig:Lx_vs_z} shows that ID830 reaches 
an absorption-corrected 0.5--2~keV luminosity of $\Lsoftx > 10^{46}$~erg~s$^{-1}$,
corresponding to an estimated bolometric luminosity of $\lbol \approx 10^{48}$~erg~s$^{-1}$.
This implies that ID830 is either an extremely massive SMBH near the maximum mass-limit of $\mbh \approx 10^{10}$~$\msun$ \citep[and thus expected Eddington luminosity is $L_\mathrm{Edd} \approx 10^{48}$~erg~s$^{-1}$,][]{net03,mcl04,ghi10,mcc11,kor13,tra14,ina16b,kin16,ich17} or an SMBH in a super-Eddington accretion phase (if $\mbh < 10^{10} \msun$).

In this paper, we present a detailed multiwavelength analysis of ID830,
using SDSS optical and Subaru/MOIRCS near-infrared spectroscopic observations covering rest-frame UV to optical bands as well as broad-band spectral energy distribution (SED) covering from 100~MHz radio to X-ray bands. 
We investigate its accretion state, jet properties, and SMBH mass, and 
discuss \rt{the implications of its unusually bright X-ray emission for SMBH growth and AGN feedback at $z\sim3$--$4$} \citep[e.g.,][]{fab12,har24}.
We adopt the following cosmological parameters throughout this paper: 
$H_0 = 68$\,km\,s$^{-1}$\,Mpc$^{-1}$, $\Omega_\mathrm{M}=0.31$, and $\Omega_\Lambda=0.69$ \citep{Planck2020}.

\section{Sample Selection}\label{sec:sample}

\begin{figure}
\begin{center}
\includegraphics[width=0.48\textwidth]{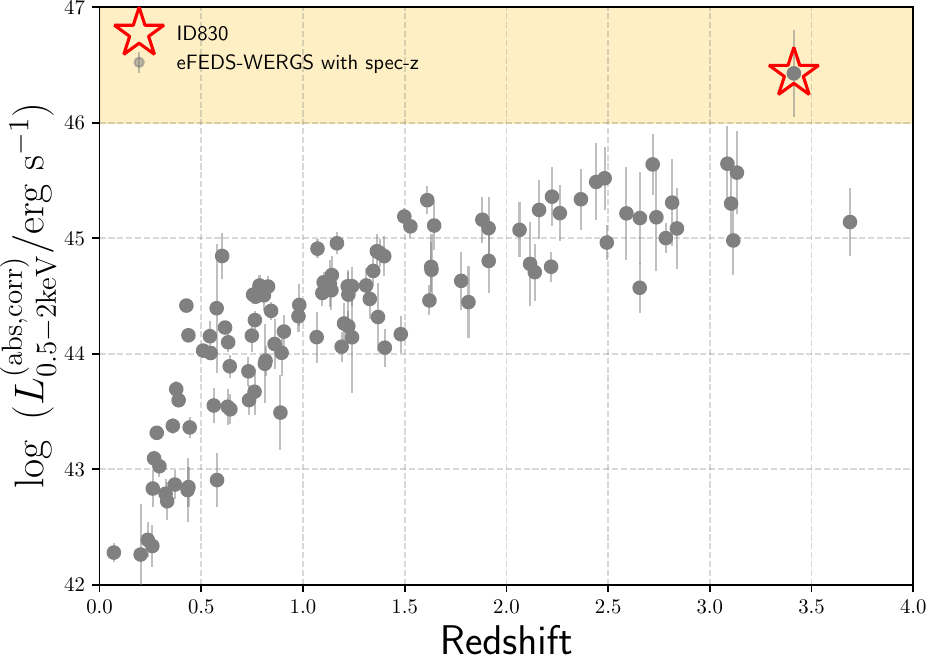}
\caption{
Absorption-corrected 0.5--2~keV Luminosity of the radio AGN sample in \cite{Ichikawa2023} as a function of redshift (gray circles). The sample shows selected 113 sources with reliable spec-$z$ and near-IR detections. ID830 is enclosed by the open red star.
The orange shaded area represents $\Lsoftx > 10^{46}$~erg~s$^{-1}$, where only ID830 resides in this region.
}\label{fig:Lx_vs_z}
\end{center}
\end{figure}

 We start the sample selection from the eROSITA detected GHz radio AGN catalog in the eFEDS field constructed by \citet{Ichikawa2023}. 
  This catalog originates from the Wide and deep Exploration of Radio Galaxies with Subaru/HSC project \citep[WERGS;][]{Yamashita2018,tob19,ich21,uch22,uch22a,yam25,zho25}, which cross-matched radio sources from the VLA/FIRST 1.4\,GHz survey with optical counterparts from the Subaru/HSC-SSP survey \citep{Miyazaki2018} and UNIONS optical survey \citep{gwy25}.
\citet{Ichikawa2023} further matched the WERGS catalog to the eFEDS X-ray source catalog \citep{sal22,liu22} using optical positions from the DESI Legacy Imaging Surveys DR8 \citep[LS8;][]{Dey2019}. The resulting eFEDS-WERGS sample consists of 393 X-ray detected radio AGNs over the 140\,deg$^2$ eFEDS footprint, with a redshift range of $0 < z < 4$. Since the parent eFEDS catalog with LS8 counterpart information contains 17\,603 sources, the eFEDS-WERGS sample corresponds to $\sim 2.2\%$ of the total population, \rt{which is smaller than the typical fraction of radio AGNs} 
\citep[$\sim10\%$,][]{kel94,Ivezić2002}. \rt{Because of the shallow FIRST sensitivity of $f_\nu > 1$~mJy, the eFEDS-WERGS sample is missing the weaker radio sources among the radio-loud quasars.}

We applied a series of selection criteria to select extremely X-ray luminous radio quasars. 
First, we restricted the sample to sources with reliable spectroscopic redshifts (spec-$z$), reducing the number to 180. Second, to enable accurate multiwavelength SED modeling, we required coverage in the near-infrared bands from the VISTA Kilo-degree Infrared Galaxy survey \citep[VIKING;][]{arn07,edg13}, which limits the survey area to 65\,deg$^2$ and reduces 113 sources with robust NIR photometry.

Finally, we selected objects with rest-frame absorption-corrected soft X-ray luminosity of $\log(L_{0.5-2\,\mathrm{keV}}/\mathrm{erg\,s^{-1}}) > 46.0$, corresponding to an estimated bolometric luminosity of $\lbol \approx 10^{48}$~erg~s$^{-1}$. This selection targets extreme sources that may be undergoing super-Eddington accretion or host the most massive SMBHs approaching the maximum-mass limit (see discussion in Section~\ref{sec:intro}). Only one source fulfilled all criteria: eFEDS J084222.9+001000 (hereafter ID830) at $z = 3.4351\pm0.0020$, as shown in Figure~\ref{fig:Lx_vs_z}.

We further investigated whether the extreme X-ray luminosity of ID830 could be attributed to relativistic beaming, as often observed as blazars. However, its steep X-ray photon index ($\Gamma = 2.43$, see discussion in Section~\ref{sec:eROSITA-spectra}) and existence of dust extinction in the rest-frame UV to optical spectrum of $A_V = 0.39$~mag (we will discuss these points in Section~\ref{sec_sub:dust_correction}) indicate that the radio jet is not aligned with the observer's line of sight. We therefore conclude that ID830 is not a blazar and select it as the target of this study.

\section{Analysis and Results}

\subsection{eROSITA X-ray Spectral analysis}\label{sec:eROSITA-spectra}

We obtain the X-ray spectrum of ID830 observed in the eFEDS field and perform an X-ray spectral analysis using \verb|Xspec| v.12.15 \citep{Arnaud1996}. 
Although \cite{liu22} already summarized a basic X-ray spectral properties for ID830, 
their spectral analysis is designed as a simple one and we thus conclude that re-analyzing the spectrum
by ourselves would provide more useful information.
We use the extracted source and background spectra for ID830 provided by \citet{liu22}. These were obtained with the eROSITA \texttt{srctool} task, which automatically determines the source and background extraction regions to maximize the signal-to-noise ratio. The source extraction radius ranges from 10~arcsec to the radius enclosing 99\% of the energy in the point spread function (PSF). For the background, an annular region was adopted, with the inner radius set where the source surface brightness falls below 5\% of the local background level, and the outer radius chosen such that the background area is 200 times larger than the source extraction area, after excluding nearby sources. We refer the reader to \citet{liu22} for further details on the observations, spectral extraction, and data reduction.
\rt{In our re-analysis, we extend the range of the photon index to better accommodate super-Eddington sources, and update the redshift using the value obtained from the recent Subaru/MOIRCS spectrum described in Section~\ref{sec:optical_spectral_analysis}.}

\begin{figure}
\begin{center}
\includegraphics[width=0.48\textwidth]{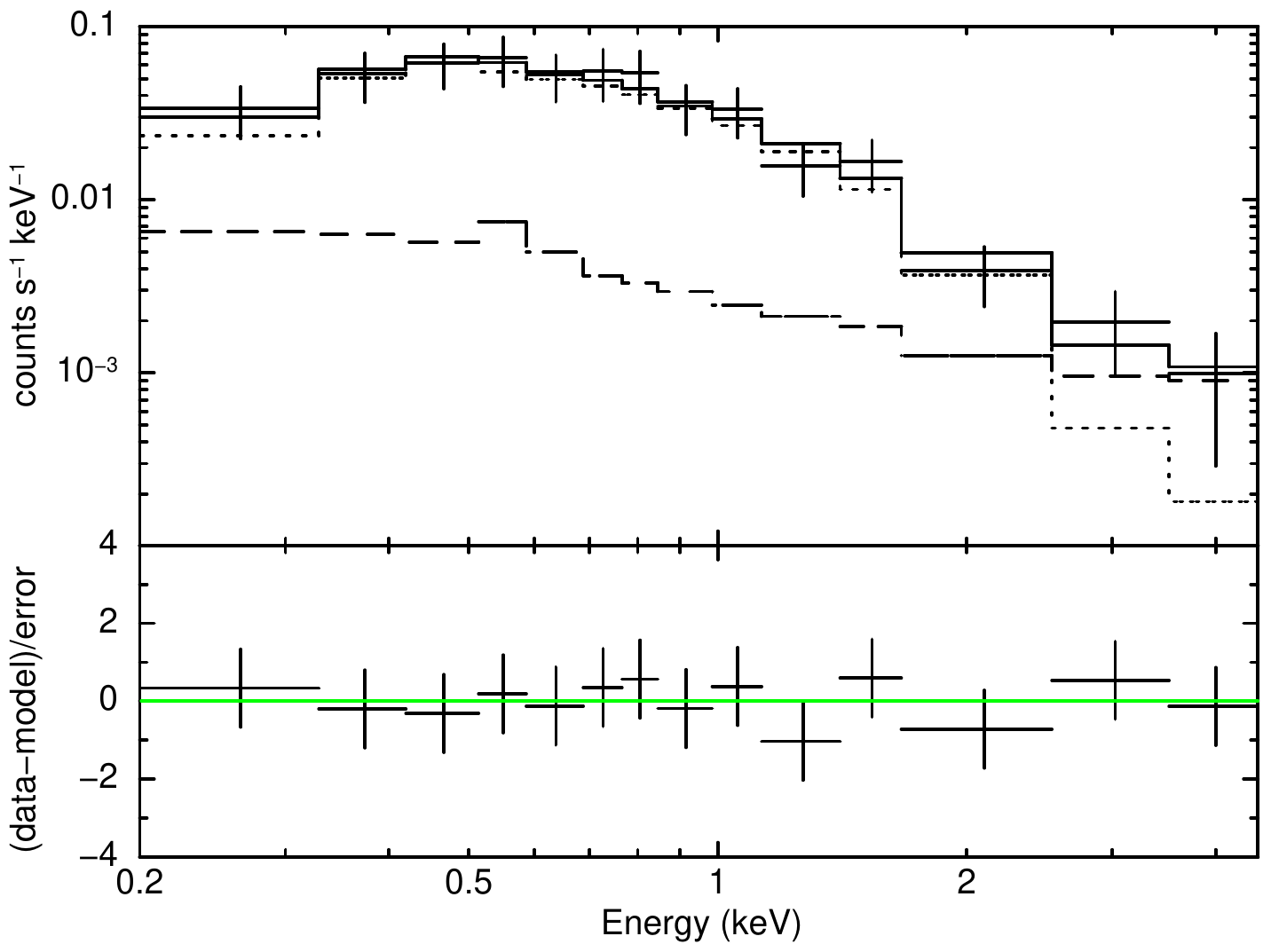}
\caption{
The X-ray spectrum in the observed frame of ID830 fitted by \texttt{Xspec}. The cross plots represent the observed data of ID830. The solid line is the best-fit spectrum, including the AGN model (dotted line) and the background model (dashed line). The fitting residuals of the best-fit model divided by the $1\sigma$ errors are shown in the lower panel.
}\label{fig:Xspec}
\end{center}
\end{figure}

Figure~\ref{fig:Xspec} shows the obtained spectrum (cross points) and the background (dashed line), which nicely covers significant detections between 0.2\,keV to 4.5\,keV (0.9\,keV to 20\,keV) in the observed (rest) frame, while the energy range above 4.5\,keV is excluded because the background is dominant in this region. 
The background spectra is fitted using the automatic method implemented in BXA \citep{Buchner2014}. In this method, principal component analysis (PCA) is performed on the background spectrum, and a linear combination of the first six principal components (PCs) is used to obtain the best-fit background model (see \citealt{liu22} for more details).

We adopt a simple model, \verb|TBabs*zTBabs*zpowerlw| in XSPEC terminology.
The model is composed of three components of the primary transmitted component
from the nucleus (\verb|zpowerlaw|) and absorption both in torus (\verb|zTBabs|, or $\NH$) and in the Galactic plane (\verb|Tbabs|, or $N_\mathrm{H,Gal}$). The Galactic absorption model is based on the Tuebingen-Boulder ISM absorption model, which calculates the Galactic X-ray absorption with the ISM abundances \citep{Wilms2000}, and it is to $N_\mathrm{H,Gal}=3.98 \times 10^{20}$\,cm$^{-2}$ at the location of ID830.
For the power-law model, we fix the redshift as $z=3.4351$. 
We allow the photon index $\Gamma$ as a free parameter within a relatively broad range
of $1< \Gamma <4$, which is wider than the typical range observed in AGN  \citep[$1.5<\Gamma<2.5$;][]{ric17,liu22}. 
This is to explore two possible AGN properties for ID830: one is as a super-Eddington AGN, which often show relatively steeper index \citep[$2<\Gamma<4$;][]{Zappacosta2023,Maithil2024} and another is as a blazar, which has much flatter photon index ($1<\Gamma<2$). 
We also adopt the $C$-statistic because of the low-count spectra (121 net counts), below the usual criterion of $\approx 200$-counts that often requires for $\chi^2$ statistics to be valid \citep{Ricci2017}
, and we fit the source and background simultaneously.

Figure~\ref{fig:Xspec} shows the best-fit model spectrum (black solid line), which is the sum of the model component (dotted line) and the background (dashed line). 
We obtain $\Gamma=2.43 \pm0.21$,
indicating that ID830 has an extremely steep X-ray spectrum (\textit{C}-statistic$=274.2$ for 315 dof).
We will discuss this point in more detail in Section~\ref{sec:X-ray_excess}.
On the other hand, our result shows that $\NH$ is not required, giving only the upper-bound of $\NH<4 \times 10^{19} $\,cm$^{-2}$.
The resultant absorption-corrected X-ray luminosities are 
$\log (\Lsoftx/\mathrm{erg}~\mathrm{s}^{-1}) = 46.20 \pm 0.12$ and
$\log (\Lhardx/\mathrm{erg}~\mathrm{s}^{-1}) = 45.99 \pm 0.17$, respectively.
A summary of the basic information including X-ray spectral fitting is tabulated in Table~\ref{tab:basic_info}.

\setlength{\topmargin}{3.0cm}
\renewcommand{\arraystretch}{1.1}
\begin{deluxetable}{lclc}
\thispagestyle{empty}
  \tablecaption{Basic information of ID830}
  \tablehead{
  \colhead{eFEDS ID} & & eFEDS J084222.9$+$001000 & Ref.\
     }
    \startdata
$z$ & & $3.4351 \pm 0.0020$ & (1) \\
$\mathrm{RA_{eFEDS}}$ & (deg) & 130.5956 & (2) \\
$\mathrm{DEC_{eFEDS}}$ & (deg) & 0.16678 & (2) \\
$\log \Lsoftx$ & ($\mathrm{erg\,s^{-1}}$) & $46.20 \pm 0.12$ & \\
$\log \Lhardx$ & ($\mathrm{erg\,s^{-1}}$) & $45.99 \pm 0.17$ & \\
$\Gamma$ & & $2.43\pm 0.21$ & \\
$\log N_\mathrm{H}$ & ($\mathrm{cm^{-2}}$) & $<19.60$ & (3) \\
SDSS ID & & SDSS J0842$+$0010 & (1) \\
$\mathrm{RA_{FIRST}}$ & (deg) & 130.5959 & (4) \\
$\mathrm{DEC_{FIRST}}$ & (deg) & 0.1674 & (4) \\
$f_\mathrm{1.4~GHz}$ & (mJy) & $16.6805 \pm 0.1523$ & (4) \\
$\log R_\mathrm{obs}$ & & $3.0215 \pm 0.0041$ & (5) \\
$\mathrm{RA_{LS8}}$ & (deg) & 130.5958 & (6) \\
$\mathrm{DEC_{LS8}}$ & (deg) & 0.1672 & (6) \\
    \enddata
    \thispagestyle{empty}
\tablenotetext{}{
Notes.--- The coordinates are at J2000. References: (1) \cite{Abdurro'uf2022}; (2) \cite{Brunner2022}; (3) \cite{liu22}; (4) \cite{Helfand2015}; (5) \cite{Ichikawa2023}; (6) \cite{Salvato2022}.
}\label{tab:basic_info}
\end{deluxetable}
\setlength{\topmargin}{0in}

\subsection{Analysis of SDSS and Subaru/MOIRCS UV Spectra and Optical SED}\label{sec:optical_spectral_analysis}

\subsubsection{Rest-frame UV Spectra and optical SED}\label{sec_sub:optical_spec_data}
To investigate the accretion disk and SMBH properties of ID830, 
we use the optical spectrum from the Sloan Digital Sky Survey \citep[SDSS; ][]{York2000}, a $J$-band spectrum obtained with Subaru/MOIRCS, as well as photometry from the VIKING $J$, $H$, $K_s$ bands and WISE W1 and W2 bands. This dataset provides a wavelength coverage of $0.3~\mu\mathrm{m}< (\lambda/\mu\mathrm{m}) < 4.6$~$\mu$m, corresponding to rest-frame wavelength range of $0.08~\mu\mathrm{m}<\lambda<1.0$~$\mu$m.

We first obtained the reduced spectra from the SDSS DR17 catalog \citep{Abdurro'uf2022}. 
The spectrum of ID830 was observed during SDSS-III survey epoch (MJD=55648, corresponding to 2011 March 28), and thus it covers the observed wavelength of
$3550\mathrm{\AA} < \lambda < 10311$\AA\ with the spectral resolution from $R=1500$ at 3800\AA\ to $2500$ at 9000\AA\ with a fiber aperture of 2 arcsec.

\begin{figure*}
\begin{center}
\includegraphics[width=0.80\textwidth]{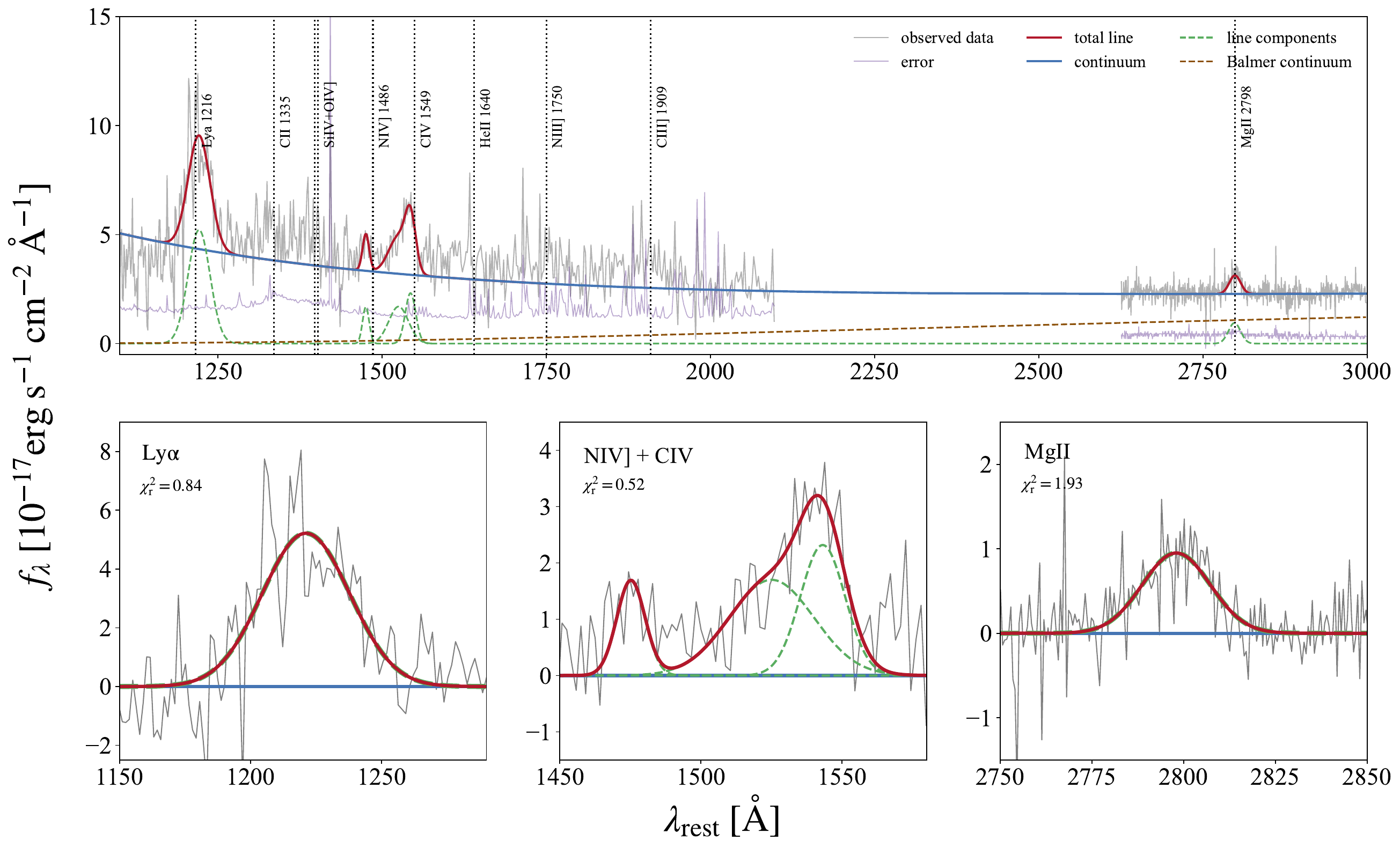}
\caption{
The optical spectra in the rest-frame of ID830 fitted by \texttt{PyQSOfit}. The spectra have been smoothed by binning the data every 5 bins for SDSS and every 3 bins for MOIRCS. The top panel shows the entire SDSS and MOIRCS spectra, and the bottom panels show the decomposed Ly$\mathrm{\alpha}$, NIV]+CIV, and MgII emission lines from left to right, respectively. In each panel, the total fitted Gaussian lines are shown in red, and the best-fit continuum is shown in blue. The green lines represent each component of the Gaussian fitting, and the brown line represents the best-fit Balmer continuum. We also write down the reduced chi-squared $\chi^2_{\mathrm{r}}$ in each bottom panel.
}\label{fig:PyQSOfit}
\end{center}
\end{figure*}

We also conducted $J$-band spectroscopic observations to cover the MgII$\lambda$2800 emission line
(S25A-0040N, PI: K.~Ichikawa).
We used the MOIRCS infrared multi-object spectrograph \citep{ich06,suz08} at the Cassegrain focus of the Subaru telescope \citep{iye04}. We observed ID830 on 2025 March 15 (HST).
The spectral configuration is MOS spectroscopic mode and we used the Lightsmyth LS\_J grism ($\lambda=1.04$--$1.3$~$\mu$m) with the spectral resolution of $R=3000$, with a slit size of 0.8~arcsec.
We follow the standard MOIRCS data reduction flow by using the MOIRCS data reduction pipeline tool MCSMDP \citep{Yoshikawa2010}, after the slight revision optimized for Python~3 environment.
We adopted a standard telescope nodding technique (ABBA pattern) with a throw of 2.5~arcsec along the slit to subtract background emission. The wavelength calibration was conducted using sky lines. We used HIP38885 as a standard star ($J=7.09$~mag) to correct for terrestrial atmospheric transmission and provide flux calibration.
The total exposure time for the target was 100~min (300~sec $\times 20$).
Figure~\ref{fig:PyQSOfit} shows the obtained J-band spectra after binning with $R=1000$ and the total SN at the continuum reaches SN$=5.1$, which is consistent with our expected values.

We further collected near-IR to mid-IR photometric points to complete the rest-frame optical SED.
We obtained 5 near-IR bands at $Z$, $Y$, $J$, $H$, and $K_S$ in 0.88, 1.02, 1.25, 1.65, and 2.15\,$\mu$m, respectively from the VISTA/VIKING catalog \citep{Kuijken2019} and also obtained 4 mid-IR bands
 at 3.4, 4.6, 12, and 22\,$\mu$m (W1, W2, W3, and W4), with the angular resolution is 6.1, 6.4, 6.5, and 12.0\,arcsec in each band. We here used unWISE photometries for W1 and W2, which provide better signal-to-noise ratio \citep{Schlafly2019,Meisner2019,Salvato2022}. All photometric information is summarized in table~\ref{tab:multiwave_info}.

\subsubsection{Dust extinction correction}\label{sec_sub:dust_correction}

\begin{figure}
\begin{center}
\includegraphics[width=0.48\textwidth]{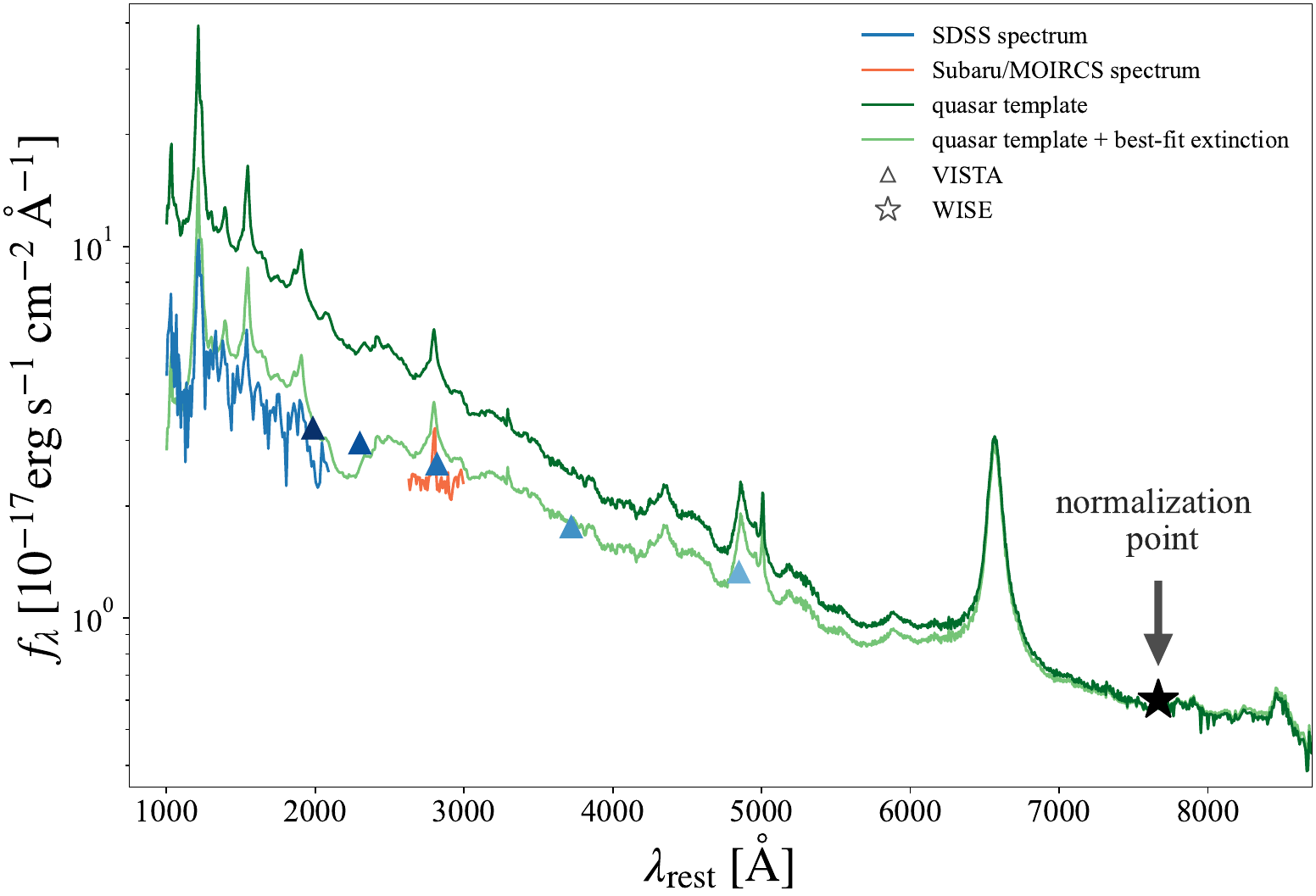}
\caption{
The result of comparing the SDSS (blue) and Subaru/MOIRCS (orange) spectra of ID830 with the normalized typical quasar template spectrum (green). We overlay the quasar template spectrum after applying the best-fit dust extinction ($A_V = 0.385$~mag, light green).
The black star represents the normalization point at the W1 band. The blue triangles denote the fluxes in each VISTA band $Z$, $Y$, $J$, $H$, and $K_S$ of ID830.
}\label{fig:normalization_point}
\end{center}
\end{figure}

Figure~\ref{fig:normalization_point} shows the combined SDSS optical and MOIRCS $J$-band spectra with VIKING+WISE W1 SED for ID830, covering the rest-frame of $\lambda=1000$--$8000$~\AA.
The composite quasar template from \cite{Selsing2016} normalized at the W1-band ($\sim$7700\,\AA\ in the rest-frame) are also overlaid to compare the spectra of ID830.
 ID830 shows the redder SED and spectra, suggesting that ID830 is a dust reddened quasar \citep[e.g.,][]{Kato2020}. 

To estimate and correct the dust extinction, we adopt the extinction curve from \cite{Gordon2016}. \cite{Gordon2016} constructed a mixture extinction curve model with the Milky Way extinction model as the $A$ component, and the SMC-like extinction model as the $B$ component (see \citealt{Gordon2016} for more details). The dust reddening law parameter $R(V)$ of this model is expressed as:
\begin{equation}
    R(V)^{-1}=f_A R_A(V)^{-1} + (1-f_A) R_B(V)^{-1},
\end{equation}
where $f_A$ is the fraction of the $A$ component and $(1-f_A)$ is the fraction of the $B$ component. We fix the fraction $f_A=1.0$ to assume the Milky Way-like extinction curve with the carbon dip at $\sim$2175\,\AA\ and the flatter curve than the Galactic shape, following the ``Maiolino''-type extinction curve for AGNs \citep{Maiolino2001a, Maiolino2001b}. Based on these parameter settings, we optimize the model to minimize the reduced chi-squared $\chi^2_\mathrm{r}$ to determine the amount of dust extinction $A_V$. As a result, we found $R_V=3.27^{+1.00}_{-0.77}$ and $A_V=0.385 \pm 0.082$~mag with $\chi^2_\mathrm{r}=0.391$.
This indicates that ID830 is a red quasar with \citep[e.g., $E(B-V)>0.1$ or corresponding to $A_V>0.33$~mag;][]{web95,gli07,urr09}. 
The light green line in Figure~\ref{fig:normalization_point} shows the dust reddened quasar template with $A_V=0.385$~mag.

\subsubsection{SDSS and MOIRCS Spectral fitting}

\setlength{\topmargin}{3.0cm}
\floattable
\renewcommand{\arraystretch}{1.1}
\begin{deluxetable*}{ccccccccccc}
\thispagestyle{empty}
  \tablecaption{Estimated black hole mass and Eddington ratio for ID830.}\label{tab:MBHandlambdaEdd}
  \tablehead{
       \colhead{Line} & \colhead{FWHM} & \colhead{$L_\mathrm{3000\,\text{\AA}}$} & 
       \colhead{$\Lhardx$} & \colhead{$L_{\mathrm{bol},3000\,\text{\AA}}$} & \colhead{$L_\mathrm{bol,2\text{-}10\,keV}$} &    
       \colhead{$\Mbh$} & \colhead{$\LEdd$} & \colhead{$\lambda_\mathrm{Edd,UV}$} & \colhead{$\lambda_\mathrm{Edd,X}$}
       \\
        & \colhead{(km\,s$^{-1}$)} & \colhead{(10$^{46}$\,erg\,s$^{-1}$)} & \colhead{(10$^{45}$\,erg\,s$^{-1}$)} & \colhead{(10$^{46}$\,erg\,s$^{-1}$)} & \colhead{(10$^{46}$\,erg\,s$^{-1}$)} & \colhead{($10^8\,\Msun$)} & \colhead{(10$^{46}$\,erg\,s$^{-1}$)}  
     }
    \startdata
\thispagestyle{empty}
MgII & $2174\pm301$ & $1.47\pm0.02$ & 
$9.74\pm3.90$ &
$7.62\pm0.31$ & $67.6\pm29.0$ & $4.40\pm0.72$ & $5.54\pm0.91$ & $1.44\pm0.24$ & $12.8\pm3.9$ \\
    \enddata
    \thispagestyle{empty}
\tablenotetext{}{
Notes.--- The estimated black hole mass, Eddington ratio, and related values obtained from the $\Mbh$ estimation. $L_\mathrm{3000\,\text{\AA}}$ represents the interpolated monochromatic luminosity at rest-frame 3000\,\text{\AA} after correcting the dust extinction. $L_\mathrm{2\text{-}10\,keV}$ also represents the absorption-corrected luminosity at rest-frame 2--10\,keV band.
The bolometric luminosity $L_\mathrm{bol,3000\,\text{\AA}}$ and $L_\mathrm{bol,2\text{-}10\,keV}$ are calculated from $L_\mathrm{3000\,\text{\AA}}$ and $L_\mathrm{2\text{-}10\,keV}$, by applying the bolometric correction BC$_\mathrm{3000\,\text{\AA}}=5.2\pm0.2$ \citep{Runnoe2012} and BC$_\mathrm{2\text{-}10\,keV}=69.3\pm12.0$ \citep{Duras2020}, respectively. 
The UV-based and X-ray-based Eddington ratios, $\lambda_\mathrm{Edd,UV}$ and $\lambda_\mathrm{Edd,X}$, are then calculated from these bolometric luminosities.}
\end{deluxetable*}
\setlength{\topmargin}{0in}

\begin{figure}
\begin{center}
\includegraphics[width=0.48\textwidth]{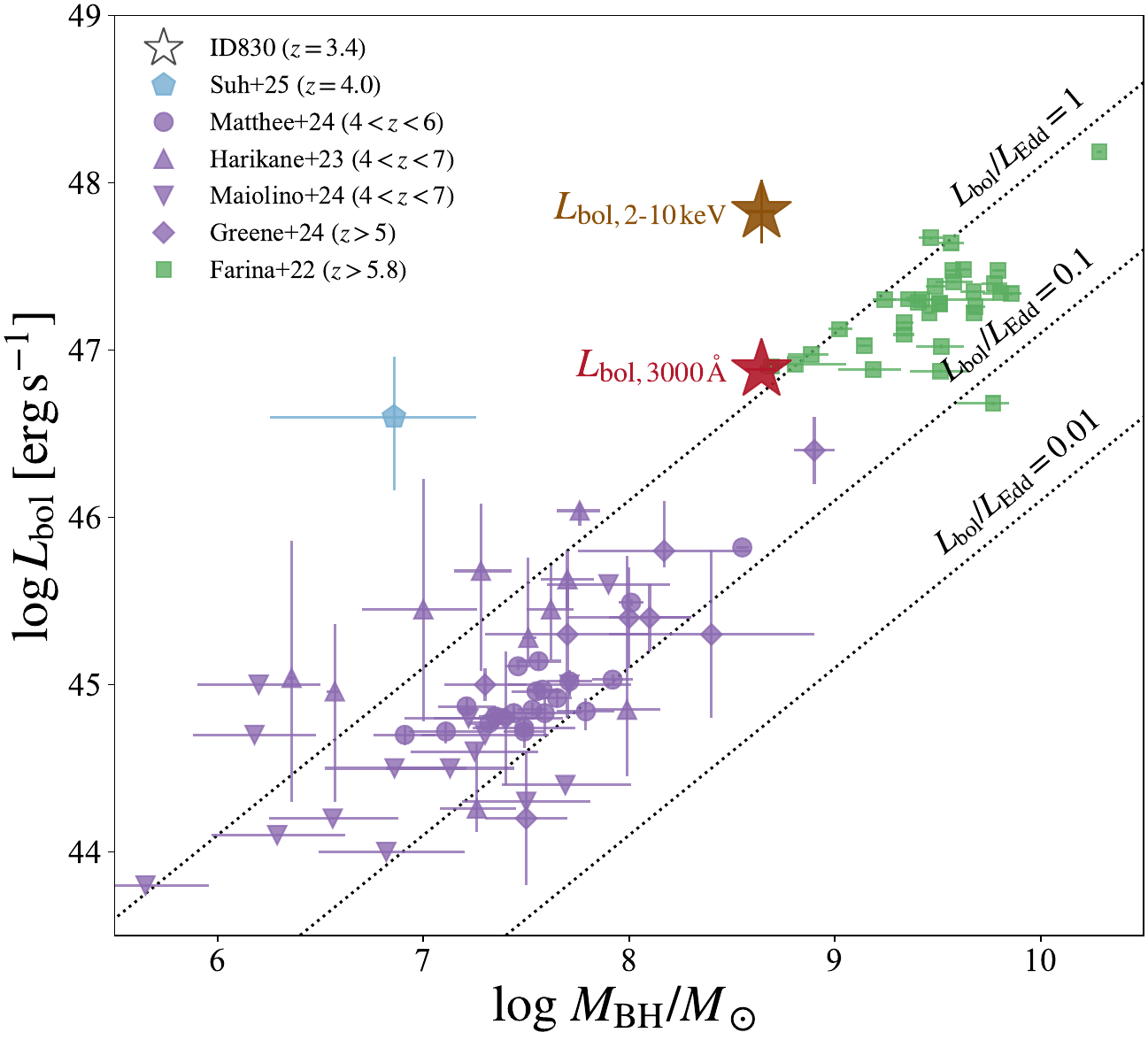}
\caption{AGN bolometric lumonosity $L_\mathrm{bol}$ vs. black hole mass $\Mbh$. For ID830, the UV-derived bolometric luminosity is shown as a red star, while the X-ray-derived bolometric luminosity is shown as a brown star. The purple markers represent JWST detected faint AGNs at $z\sim4$--7 \citep{Harikane2023,Greene2024,Maiolino2024,Matthee2024}. and the green markers represent UV-selected quasars at $z>5.8$ \citep{Farina2022}. The JWST-detected super-Eddington AGN, which is bright in X-rays, is also shown as a light blue marker \citep{suh25}.
}\label{fig:LbolvsMBH}
\end{center}
\end{figure}

We then perform a simultaneous spectral fitting of the SDSS and MOIRCS spectra using \verb|PyQSOfit| \citep{Guo2018} to probe the properties of the SMBH accretions.
\verb|PyQSOfit| is a Python code developed to investigate the spectral properties of SDSS quasars. It takes input parameters such as observed wavelengths, flux densities, errors, and redshifts from catalogs, and performs a $\chi^2$-based spectral fitting in the rest-frame wavelength. 
This code also corrects the Galactic extinction automatically based on the dust map from \cite{Schlegel1998}
and dust extinction value of $\av=0.39$~mag obtained in Section~\ref{sec_sub:dust_correction}.
We fixed $z=3.4351$ and $\av=0.39$~mag.

We first obtain the continuum fitting results as a combination of the power-law and Balmer continuum models. 
In the \verb|PyQSOfit| fitting, we use the Balmer continuum template of \cite{Dietrich2002}. 
Although Fe II emission is a significant feature of quasar spectra, forming a pseudo-continuum through blended multiplets in the rest-frame UV range of $\sim$2000–4000~\AA, we do not include the component in our fits. This is because the available spectral coverage from SDSS and MOIRCS (1000–2300~\AA\ and 2600–3000~\AA, respectively) lacks the continuous wavelength range necessary for a reliable Fe~II continuum fit.

After subtracting the best-fit continuum model, we analyze the broad-line profiles of the three emission lines: Ly$\mathrm{\alpha}\lambda1216$\,\AA, CIV$\lambda1549$\,\AA\ and MgII$\lambda2798$\,\AA, using a Gaussian fitting. The broad-line components of these emission lines are defined as having a full width at half-maximum (FWHM) $\geq$ 1200\,km\,s$^{-1}$, and the peak wavelength range around the broad emission lines is set to 1150--1300\,\AA\ for Ly$\mathrm{\alpha}$, 1530--1550\,\AA\ for CIV, and 2600--3000\,\AA\ for MgII. 

The CIV line is often blueshifted as it is emitted from the outflowing gas driven by AGN \citep[e.g.,][]{Coatman2017}. To account for this, we fit the CIV emission with two line profiles to represent the systematic and outflow components.
The uncertainties on both the CIV velocity dispersion and the outflow velocity are conservatively set to 20\%. 
We simultaneously fit the NIV]$\lambda\lambda1486$~\AA\ emission line, which is detected in only $\sim$5\% of the quasar population \citep[the so-called nitrogen-loud quasars;][]{osm80,bal03}. Its detection in ID830 is reasonable, given that more than 50\% of nitrogen-loud quasars are radio-loud \citep{Jiang2008}.

For the fitting, the peak wavelength range is set to 1515--1540~\AA\ for the CIV outflow component and 1450--1500~\AA\ for NIV], with an upper limit on the velocity dispersion corresponding to $\sigma \leq 15$~\AA\ for the CIV outflow component and $\sigma \leq 5$~\AA\ for NIV]. Figure~\ref{fig:PyQSOfit} shows the results of the spectral fitting. We find clear broad emission lines of Ly$\alpha$ (FWHM$_\mathrm{Ly\alpha} = 9642 \pm 915$~km~s$^{-1}$), CIV (FWHM$_\mathrm{CIV} = 3624 \pm 729$~km~s$^{-1}$), and MgII (FWHM$_\mathrm{MgII} = 2174 \pm 301$~km~s$^{-1}$). The outflow velocity of CIV is $v_\mathrm{outflow}=4645 \pm 929$~km~s$^{-1}$.

Based on the fitting results, we estimate the black hole mass and the Eddington ratio of ID830. 
We use the single-epoch virial mass estimation \citep{Shen2011}:
\begin{equation}
    \log \bigg(\frac{M_{\mathrm{BH, vir}}}{\Msun} \bigg) = a + b\log \bigg(\frac{\lambda L_{\lambda}}{10^{44} ~ \mathrm{erg~s^{-1}}}  \bigg) \\
    + 2\log \bigg(
    \frac{\mathrm{FWHM}}{\mathrm{km~s^{-1}}} \bigg),
\end{equation}
where the coefficients $a$ and $b$ are parameters, which vary depending on the type of the emission lines.
The CIV-based mass estimate is known to be biased compared to other single-epoch mass estimations such as those based on H$\alpha$, H$\beta$, and MgII, due to outflows from AGNs that can cause the CIV emission line to be blueshifted \citep[e.g.,][]{Coatman2017}. Therefore, we calculate the MgII-based $\Mbh$ using the estimated FWHM$_\mathrm{MgII}$ and the continuum luminosities $L_\mathrm{3000\,\text{\AA}} = (1.47 \pm 0.02) \times10^{46}$\,erg\,s$^{-1}$ (see Table~2), taking into account the dust extinction correction discussed in Section~\ref{sec_sub:dust_correction}.

For the MgII-based estimate, we use the three sets of the parameters from \cite{McLure2004}, \cite{Vestergaard2009}, and \cite{Shen2011}, respectively:
\begin{equation}
    (a, b)=(0.505,~0.62), 
\end{equation}
\begin{equation}
    (a, b)=(0.860,~0.50), 
\end{equation}
\begin{equation}
    (a, b)=(0.740,~0.62).  
\end{equation}
The three estimated masses are combined by the Monte Carlo simulation. Each measurement is modeled as a Gaussian distribution with its estimated mean and $1 \sigma$ uncertainty. A total of $N=10^6$ samples are generated randomly from these distributions, and for each trial the sampled values are averaged to construct a combined black hole mass distribution. The mean and standard deviation are adopted as a representative black hole mass and its uncertainty, respectively.
The calculated BH mass is determined to be $\Mbh=(4.40\pm0.72) \times 10^8\,\Msun$. Utilizing the bolometric luminosity derived from the absorption-corrected UV luminosity of $L_\mathrm{bol,3000\,\text{\AA}}=7.62\times10^{46}$\,erg\,s$^{-1}$, the UV-based Eddington ratio of ID830 is determined to be $\lambda_\mathrm{Edd,UV}=1.44\pm0.24$. 
Based on this result, we conclude that ID830 is a super-Eddington quasar with $\lambdaedd >1$.
 
The bolometric luminosity can also be estimated independently from the X-ray band, using the absorption-corrected rest-frame 2--10\,keV luminosity $\Lhardx$ and the bolometric correction of \cite{Duras2020}, which yields $L_\mathrm{bol,2\text{-}10\,keV}=6.76\times10^{47}$\,erg\,s$^{-1}$. 
Figure~\ref{fig:LbolvsMBH} presents the relation between $\Mbh$ and $L_\mathrm{bol}$, including both the UV-based and X-ray-based bolometric luminosities of ID830. For comparison, we also show low-mass black holes at $z\sim4$--7 recently discovered with JWST, as well as UV-selected quasars at $z>5.8$. The AGN reported by \cite{suh25}, low-mass, super-Eddington source at $z=3.965$ discovered with JWST, is notably bright in X-ray emission, similar to ID830. For ID830, it is clear that $L_\mathrm{bol,3000\,\text{\AA}}$ and $L_\mathrm{bol,2\text{-}10\,keV}$ are inconsistent, implying that the UV luminosity is unusually faint, or alternatively, the X-ray luminosity is unusually bright. Based on $L_\mathrm{bol,2\text{-}10\,keV}$, we infer the Eddington ratio of $\lambda_\mathrm{Edd,X}=12.8\pm3.9$, which is $\sim 1$\,dex higher than the UV-based $\lambda_\mathrm{Edd,UV}$. In Section~\ref{sec:alphaOX}, we discuss this discrepancy with respect to the accretion disk properties.

\subsection{Radio Properties}\label{sec:radioproperties}
\begin{figure}
\begin{center}
\includegraphics[width=0.48\textwidth]{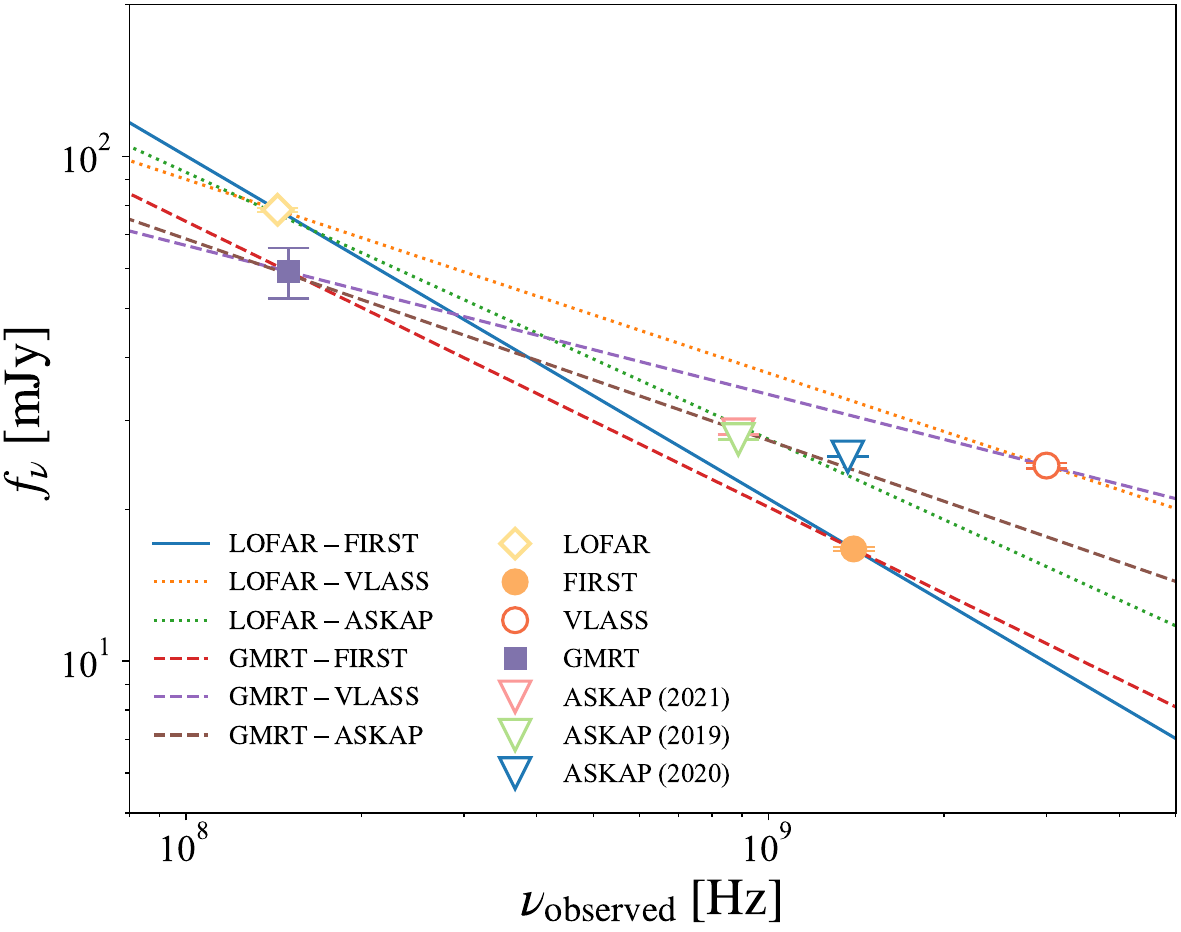}
\caption{
The radio SED in the observed frame of ID830. The colored markers represent the radio measurements. The colored lines also represent the slopes of the spectrum between each pair of plots. In particular, the blue solid line shows the LOFAR--FIRST slope adopted in this paper.
}\label{fig:radio_SED}
\end{center}
\end{figure}

\begin{figure*}
\begin{center}
\includegraphics[width=0.8\textwidth]{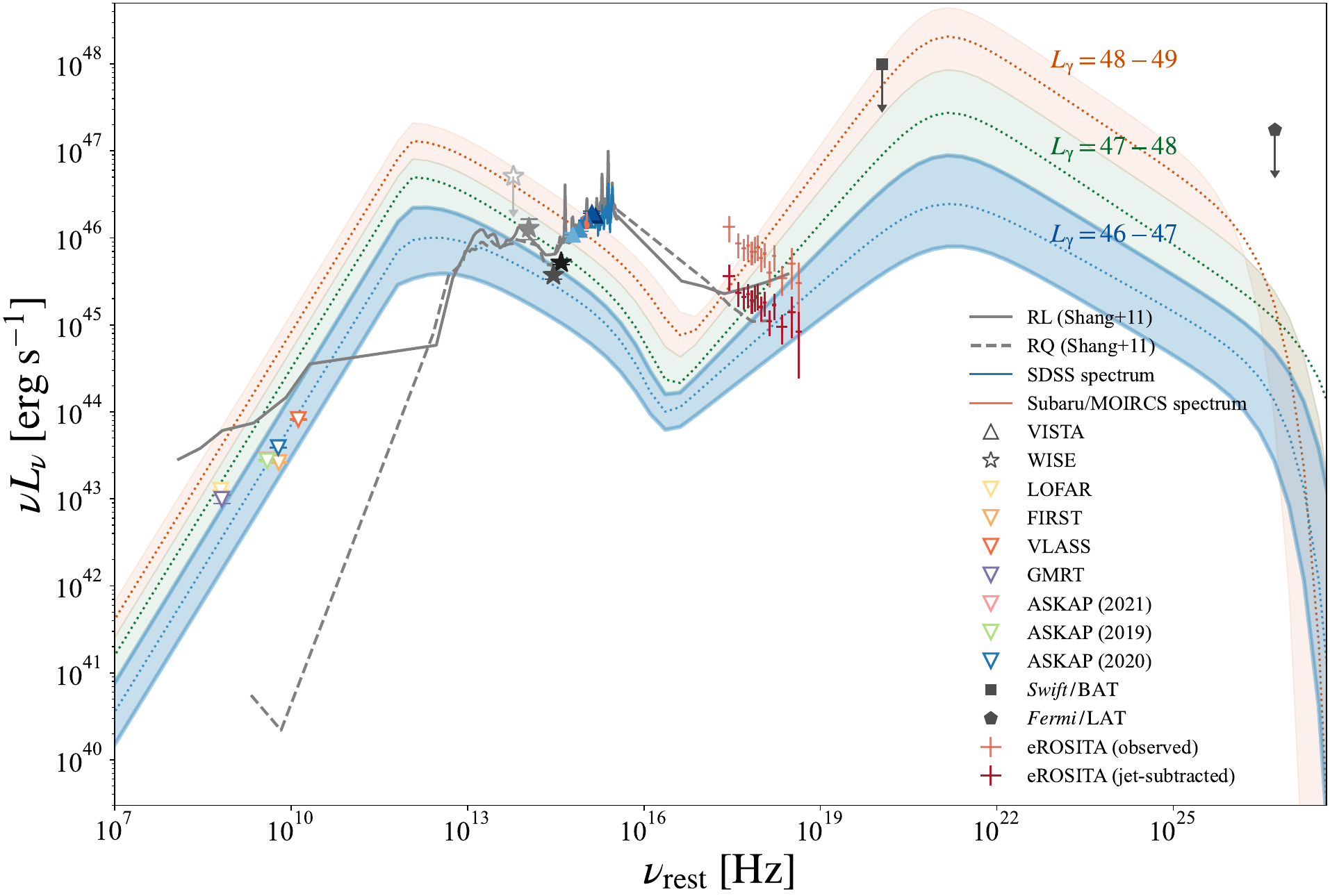}
\caption{
The broad-band SED of ID830, along with the SED templates of radio-loud and radio-quiet quasars \citep{Shang2011} and the blazar sequence \citep{Ghisellini2017}. The blue, green, and orange shaded regions represent the individual bins of the blazar sequence. The light red plots are the \rt{observed} X-ray spectrum of ID830. The dark red plots are the \rt{jet-subtracted} X-ray spectrum, after removing the jet contamination. The upper limits are indicated by arrows.
}\label{fig:multiwave_SED}
\end{center}
\end{figure*}

We obtained multi-band radio flux densities from various epochs, ranging from 1990s to 2020.
We collected flux densities from the following catalogs: LOw Frequency ARray 144~MHz \citep[LOFAR;][]{vanHaarlem2013}, the Very Large Array Sky Survey 3GHz \citep[VLASS;][]{Lacy2020}, the Giant Metrewave Radio Telescope (GMRT) 150\,MHz All-sky Radio Survey \citep[TGSS-ADR1;][]{Intema2017}, and the Australian Square Kilometre Array Pathfinder 0.89 and 1.37~GHz \citep[ASKAP;][]{Johnston2008,Hotan2021} Survey of GAMA-09 + X-Ray (SWAG-X; Moss et al., in prep.) and all radio flux densities are summarized in Table~\ref{tab:multiwave_info}.

Figure~\ref{fig:radio_SED} shows the radio SED of ID830, for which we determine a slope index of $\alpha = 0.68$, adopting the slope between the LOFAR 144~MHz and FIRST~1.4~GHz data points as representative points and by assuming a power-law form $f_\nu \propto \nu^{-\alpha}$.
On the other hand, the slope index gets slightly shallower of $\alpha=0.38$ if we use the LOFAR 144~MHz and VLASS 3~GHz as representative points.

All radio images show point source features, suggesting that obvious lobe emissions are not detected in any radio images. The most stringent upper-bound is given by VLASS~3GHz 2.5~arcsec image, which provides the upper-limit of the jet-lobe size of 18.8~kpc. Since this size is smaller than typical FR II/I jet size of $\gtrsim 100$~kpc \citep[e.g.,][]{an12}, \rt{ID830 might host a young, powerful radio jet.} We will discuss its jet power in Section~\ref{sec:jetproduction}.

\subsection{Broad-band Spectral Energy Distribution}\label{sec:multiwaveSED}
\setlength{\topmargin}{3.0cm}
\floattable
\renewcommand{\arraystretch}{1.1}
\begin{deluxetable*}{cccccc}
\thispagestyle{empty}
  \tablecaption{Summary of ID830 multiwavelength information.}
  \tablehead{
       \colhead{Catalog} & \colhead{log $\nu_{\mathrm{rest}}$} & \colhead{$\lambda_{\mathrm{rest}}$} & \colhead{$f_{\nu,\mathrm{rest}}$} & \colhead{log $\nu L_{\nu}$} & \colhead{Ref.} \
       \\
       \ & \colhead{(Hz)} & \colhead{($\mu$m)} & \colhead{(mJy)} & \colhead{(erg\,s$^{-1}$)} & \ 
     }
    \startdata
\thispagestyle{empty}
VISTA $Z$ & 15.18 & 0.20 & $0.0043 \pm 0.00004$ & $45.86 \pm 0.004$ & (1) \\
VISTA $Y$ & 15.11 & 0.23 & $0.0052 \pm 0.0001$ & $45.88 \pm 0.01$ & (1) \\
VISTA $J$ & 15.02 & 0.28 & $0.0069 \pm 0.0001$ & $45.91 \pm 0.008$ & (1) \\
VISTA $H$ & 14.90 & 0.37 & $0.0081 \pm 0.0003$ & $45.86 \pm 0.02$ & (1) \\
VISTA $K_S$ & 14.79 & 0.48 & $0.0104 \pm 0.0005$ & $45.86 \pm 0.02$ & (1) \\
WISE W1 & 14.59 & 0.77 & $0.0096 \pm 0.0005$ & $45.62 \pm 0.02$ & (2) \\
WISE W2 & 14.46 & 1.04 & $0.0102 \pm 0.0011$ & $45.52 \pm 0.05$ & (2) \\
WISE W3 & 14.04 & 2.71 & $0.1016 \pm 0.0272$ & $46.10 \pm 0.12$ & (2) \\
WISE W4 & 13.78 & 5.00 & $< 0.7549$ & $<46.71$ & (3) \\
VLASS & 10.12 & -- & $5.50 \pm 0.073$ & $43.91 \pm 0.01$ & (4) \\
FIRST & 9.79 & -- & $3.76 \pm 0.03$ & $43.42 \pm 0.004$ & (5) \\
GMRT & 8.82 & -- & $13.33 \pm 1.53$ & $43.00 \pm 0.05$ & (6) \\
LOFAR & 8.80 & -- & $17.67 \pm 0.18$ & $43.10 \pm 0.004$ & (7) \\
ASKAP (2020) & 9.78 & -- & $5.73 \pm 0.02$ & $43.59 \pm 0.001$ & (8) \\
ASKAP (2019) & 9.60 & -- & $6.20 \pm 0.02$ & $43.44 \pm 0.001$ & (8) \\
ASKAP (2021) & 9.60 & -- & $6.34 \pm 0.02$ & $43.45 \pm 0.001$ & (8) \\
\textit{Swift}\ BAT & 20.05 & -- & -- & $<48.00$ & (9) \\
\textit{Fermi}\ LAT & 26.73 & -- & -- & $<47.24$ & (10) \\
    \enddata
    \thispagestyle{empty}
\tablenotetext{}{
Notes.--- Reference: (1) \cite{Kuijken2019}; (2) \cite{Meisner2019}; (3) \cite{Cutri2013}; (4) \cite{Lacy2020}; (5) \cite{Helfand2015}; (6) \cite{Intema2017}; (7) \cite{vanHaarlem2013}; (8) Moss et al., in prep.; (9) \cite{Oh2018}; (10) \cite{Fermi2020}.
}\label{tab:multiwave_info}
\end{deluxetable*}
\setlength{\topmargin}{0in}

Figure~\ref{fig:multiwave_SED} shows the broad-band SED.
In addition to already discussed data points in radio, IR, optical, and X-ray spectra, we further added
upper-limits from the \textit{Swift}/BAT (Burst Alert Telescope) 157-month catalog \citep{Lien2025} with a $5\sigma$ sensitivity limit of $f_\mathrm{14\text{-}195keV,lim}=8.83 \times 10^{-12}~\mathrm{erg\,s^{-1}\,cm^{-2}}$, and from the \textit{Fermi}/LAT (Large Area Telescope) 4th catalog \citep{Fermi2020} with a Test Statistic $TS>25$ ($\sim 4\sigma$) sensitivity limit of $f_\mathrm{50MeV-1TeV,lim}=1.56 \times 10^{-12}~\mathrm{erg\,s^{-1}\,cm^{-2}}$ at the target position. 
A summary of the multiwavelength IR, radio, and $\gamma$-ray information is presented in Table~\ref{tab:multiwave_info}.

Figure~\ref{fig:multiwave_SED} shows both the original X-ray spectrum (orange crosses) and corrected one after subtracting jet-linked X-ray emission (red crossed, see discussion in Section~\ref{sec:X-ray_excess}). To investigate the properties of ID830 as a radio-loud quasar, we also overlay the SED templates of radio-loud and radio-quiet quasars \citep{Shang2011}, normalized at the VISTA $Z$ band flux. As shown in Figure~\ref{fig:multiwave_SED}, the radio and optical plots of ID830 follow the radio-loud template well. A peak in the optical and near-IR bands indicates emission from the accretion disk. The excess seen in WISE W3 band ($\lambda=12$~$\mu$m, which corresponds to $\lambda_\mathrm{rest} \approx 3$~$\mu$m) is likely due to IR emission from the hot dust emission heated by AGN. 
However, the original X-ray plots obviously exceed the emission consistent with other bands, similar to the X-ray excess found in Figure~\ref{fig:alphaOX}. The plots also represent the very steep spectrum.

In addition, we overlay the blazar sequence of \cite{Ghisellini2017}, which includes only inverse Compton and synchrotron radiation. This blazar sequence was constructed using \rt{flat-spectrum radio quasars (FSRQs)} from the \textit{Fermi}/LAT third catalog, and classified into three luminosity bins: $46 < \log L_\mathrm{\gamma} < 47$, $47 < \log L_\mathrm{\gamma} < 48$, and $48 < \log L_\mathrm{\gamma} < 49$. Based on radio luminosities, the SED of ID830 is most consistent with the $46 < \log L_\mathrm{\gamma} < 47$ bin. Thus, although ID830 is unlikely to be a blazar, it nevertheless exhibits a higher X-ray luminosity than those in the blazar sequence.

\subsection{$\alphaox$ vs. $\LuvalphaOX$}\label{sec:alphaOX}
\begin{figure}
\begin{center}
\includegraphics[width=0.48\textwidth]{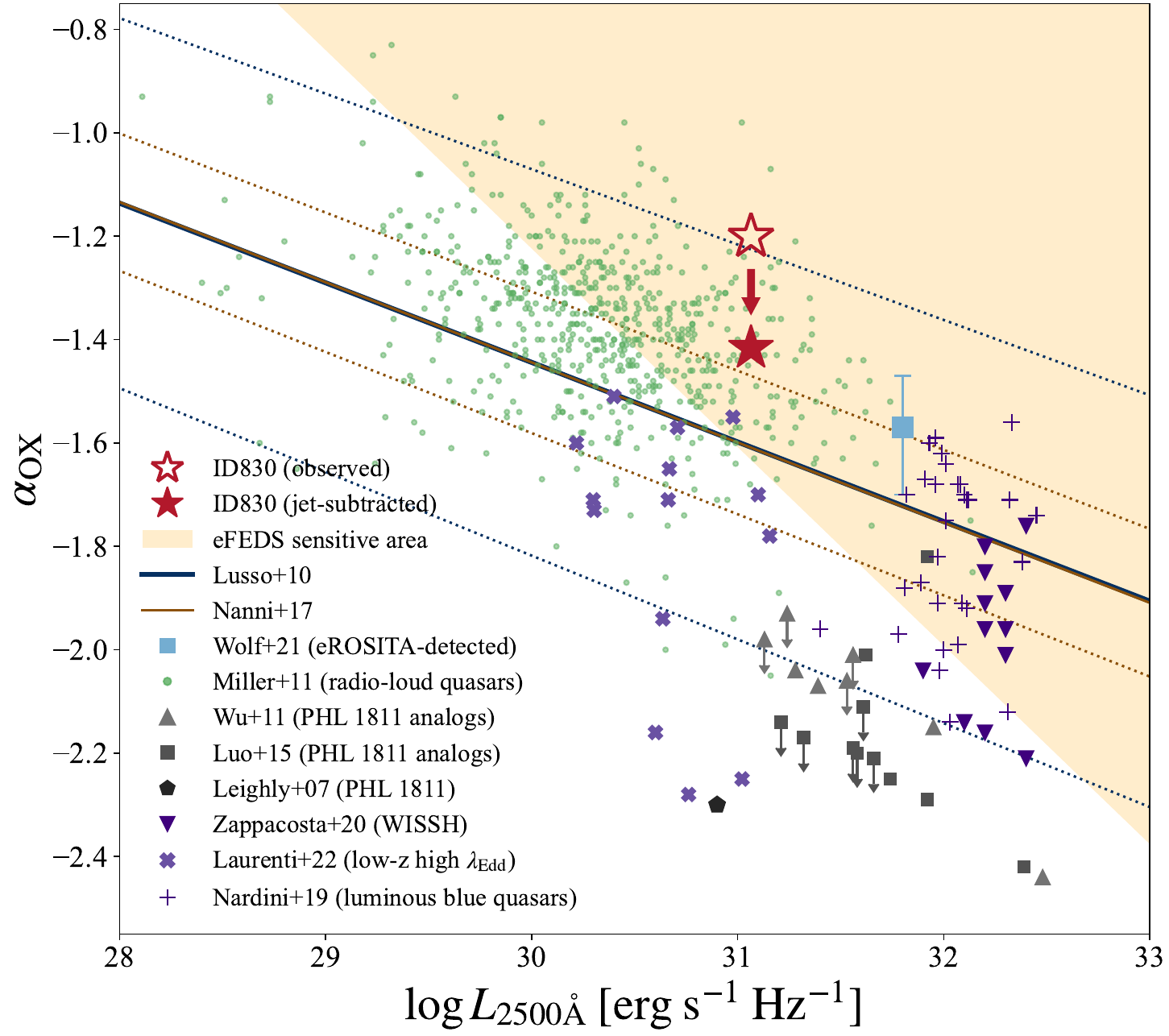}
\caption{
$\alphaox$ vs. UV monochromatic luminosity $\LuvalphaOX$ for ID830. The open star represents the \rt{observed} $\alphaox$ of ID830. The filled star represents the \rt{jet-subtracted} $\alphaox$, after removing the jet contamination. \rt{Both $\alphaox$ points are higher than} the best-fit relation of \cite{Lusso2010} (blue line) and \cite{Nanni2017} (brown line). \rt{The dotted lines are the $1 \sigma$ upper and lower bounds of the two best-fit relations.} The square plot is a eROSITA-detected quasar \citep{Wolf2021}, and black plots are PHL 1811 and its analogs \citep{Leighly2007,Wu2011,Luo2015}. Other X-ray weak sources are also shown as purple inverted triangles \citep{Zappacosta2020}, bold cross markers \citep{Laurenti2022}, and plus markers \citep{Nardini2019}.
Radio-loud quasars with X-ray excesses detected by \textit{Chandra}, \textit{XMM-Newton}, and ROSAT \citep{Miller2011} are shown as green circles. The orange shaded region shows the eFEDS sensitive area.
}\label{fig:alphaOX}
\end{center}
\end{figure}

Our results show that, while ID830 is a radio-loud quasar reaching super-Eddington accretion,
its X-ray emission shows two interesting signature; one is its significant X-ray excess compared to other quasars,
and the other is that its X-ray spectrum shows the steep photon index with $\Gamma=2.43$.

A popular method for investigating the X-ray properties of AGNs is to compare them with their UV luminosities. The ratio of UV to X-ray luminosities is known as the $\alphaox$ parameter \citep{Tananbaum1979}, which is defined as:
\begin{equation}\label{eq_alphaOX}
    \alphaox=0.384 \times \log \bigg( \frac{L_{\mathrm{2\,keV}}}{\LuvalphaOX}\bigg),
\end{equation}
where $L_\mathrm{2\,keV}$ and $\LuvalphaOX$ are the rest-frame monochromatic luminosities with the unit of $\mathrm{erg\,s^{-1}\,Hz^{-1}}$ at 2\,keV and 2500\,\AA, respectively. $\alphaox$ can be a clear tracer of the balance between accretion disk and hot corona emissions. We interpolate $\LuvalphaOX$ from the observed fluxes in each VISTA band, and $L_\mathrm{2\,keV}$ is calculated from the 0.5--2\,keV rest-frame luminosity, following \cite{Wolf2021}. $L_\mathrm{2\,keV}$ can be converted from $\Lsoftx$ depending on the photon index $\Gamma$:
\begin{equation}
    L_{\mathrm{2\,keV}}=\frac{\Lsoftx}{\int_{\nu_{\mathrm{0.5\,keV}}}^{\nu_{\mathrm{2\,keV}}}\nu^{1-\Gamma}d\nu}\nu_{2\,\mathrm{keV}}^{1-\Gamma}.
\end{equation}

The anti-correlation between $\alphaox$ and $\LuvalphaOX$ for UV bright quasars is observationally reported by several studies. \cite{Lusso2010} established the relation of
\begin{equation}
    \alphaox=(-0.154 \pm 0.010)\log \LuvalphaOX+(3.176\pm 0.223),
\end{equation}
from XMM-COSMOS detected type 1 AGN samples over wide range of redshift ($0.04<z<4.25$). \cite{Nanni2017} also found a similar relation of
\begin{equation}
    \alphaox=(-0.155\pm 0.003)\log \LuvalphaOX+(3.206\pm 0.103),
\end{equation}
from X-ray and optical selected quasars including their high-$z$ ($z>5.5$) samples observed with \textit{Chandra}, \textit{XMM-Newton} and \textit{Swift}-XRT. In addition, there is no evidence for a significant evolution of $\alphaox$--$\LuvalphaOX$ relation with redshift, suggesting that the structure of the accretion disk and the hot corona does not change substantially over cosmic time \citep[e.g.,][]{Vito2019}. 

Figure~\ref{fig:alphaOX} shows the $\alphaox$--$\LuvalphaOX$ slopes and the $\alphaox$ of ID830 derived in this work. The open marker indicates the original value of $\alphaox=-1.201\pm \rt{0.067}$ calculated using Equation~\ref{eq_alphaOX}
and values for ID830 of $L_{\mathrm{2\,keV}}=10^{27.9\pm0.14}$~erg\,s$^{-1}$\,Hz$^{-1}$ and $\LuvalphaOX=10^{31.1\pm0.11}$~erg\,s$^{-1}$\,Hz$^{-1}$. As shown in Figure~\ref{fig:alphaOX}, ID830 has the X-ray excess compared to the best-fit slopes. This result suggests that the inner structure of ID830 would be more efficient in producing X-ray emission compared to standard disk model AGNs with similar UV luminosities. 
Furthermore, considering the super-Eddington characteristic of ID830, the X-ray excess is particularly prominent because previous high-$\lambdaedd$ AGN samples typically exhibit X-ray weakness, contrary to ID830.

PHL 1811 analogs \citep[e.g.,][]{Wu2011,Luo2015} shown in Figure~\ref{fig:alphaOX}, which have unusual UV properties similar to PHL 1811 \citep{Leighly2007}, are a type of BAL (broad absorption line) quasars \citep{Lynds1967}. Most of them have high, or even super-Eddington accretion rates. They are also typically weak in X-rays, suggesting that substantial mass accretion tends to reduce the X-ray emission in comparison to the UV emission. We also show other high $\lambdaedd$ sources in Figure~\ref{fig:alphaOX}, such as WISE/SDSS-selected Hyper-luminous (WISSH) quasars at $2<z<4$ \citep{Zappacosta2020}, low-$z$ high-$\lambdaedd$ quasars at $0.4<z \le 0.75$ observed with \textit{XMM}-\textit{Newton} \citep{Laurenti2022}, and X-ray luminous, non-BAL blue quasars at $3.0<z<3.3$ observed with \textit{XMM}-\textit{Newton} \citep{Nardini2019}. All of these sources have high Eddington ratios and exhibit X-ray weakness, once again suggesting that higher accretion rates suppress X-ray production.

Based on this, we conclude that ID830 obviously exhibits a significant X-ray excess and that some factor may contribute to the excess. The most reasonable cause to explain this behavior is the jet-linked components of X-rays. Radio-loud quasars have been found to show the X-ray excess due to jet-linked X-ray emission \citep{Miller2011}, and ID830 appears to be an example of such quasars. The possibility of contamination from jet-linked components will be discussed in Section~\ref{sec:X-ray_excess}.

\section{Discussions}\label{sec:discussion}
\rt{We here discuss the implications of the unique properties of ID830 for its X-ray and radio emission mechanisms, as well as for the number density of radio-loud AGNs. The possible origin of the X-ray excess is discussed in Section~\ref{sec:X-ray_excess}, and the jet production in the super-Eddington phase is discussed in Section~\ref{sec:jetproduction}. In Section~\ref{sec_sub:LF}, we also discuss the radio-AGN fraction based on the X-ray and UV luminosity functions.}

\subsection{What Causes the X-ray Excess?}\label{sec:X-ray_excess}
In this section, we discuss the causes of the X-ray excess that has been repeatedly mentioned above. The possible contribution from the jet is described in Section~\ref{sec:jetcontamination}, and that from the disk-linked is described in Section~\ref{sec:diskcontamination}. We also discuss the transition of $\alphaox$ in the evolution of a post-outburst source in Section~\ref{sec:alphaoxtransition}.

\subsubsection{Contamination from the jet component}\label{sec:jetcontamination}
One possible origin of the X-ray excess is contamination from jet-linked components. Previous studies have shown that radio-loud quasars tend to be brighter in X-rays than their radio-quiet counterparts due to their jet emission \citep[e.g.,][]{Zamorani1981,Miller2011}. The jet-linked X-ray emission is likely to arise from inverse Compton scattering by high-energy electrons in the jet. The seed photons may originate from the jet itself (synchrotron self-Compton; SSC) or from the central regions of the AGN, such as the accretion disk and dust torus (external-Compton; EC), although their exact origin remains unclear. 

Inverse Compton scattering of the cosmic microwave background photons (IC/CMB; e.g., \citealt{Tavecchio2000,cel01}) is also one of the possible jet-linked X-ray emissions.
For high-z sources at $z > 6$, the IC/CMB contribution can be a possible origin \citep[e.g.,][]{Medvedev2020}.
However, since the number of CMB photons decreases with decreasing redshift and the contribution of IC/CMB is proportional to $(1+z)^4$ \citep{Schwartz2002}, we conclude that IC/CMB is not particularly dominant in ID830, which is at $z=3.4351$.
Indeed, we verify that the corresponding shift is $\Delta \alphaox \approx -0.05$ based on the X-ray enhancement factor at $z=3.4$ from \cite{Zhu2019}. 

We here estimate the jet-linked X-ray emission and isolate the corona-linked X-ray emission to investigate the intrinsic X-ray properties without the jet contribution. \cite{Miller2011} presented the X-ray properties of radio-loud quasars (including radio-intermediate quasars) selected from \textit{Chandra}, \textit{XMM}-\textit{Newton}, and ROSAT archival data with optical data (e.g., SDSS) and radio data (e.g., VLA/FIRST 1.4~GHz). They measured the excess X-ray luminosity of radio-loud quasars (RLQs) compared to corresponding radio-quiet quasars (RQQs), and constructed a best-fit model as a function of radio-loudness:
\begin{equation}
    \log \bigg(\frac{L_{\mathrm{2keV,RLQ}}}{L_{\mathrm{2keV,RQQ}}} \bigg) = (-0.354 \pm 0.050)+(0.352 \pm 0.039)\,\log R^*,
\end{equation}
where $L_{\mathrm{2keV,RLQ}}$ and $L_{\mathrm{2keV,RQQ}}$ are the X-ray luminosities of RLQs and RQQs at rest-frame 2\,keV, respectively, and $\log R^*$ is the radio-loudness defined by
$\log R^* = \log\, (L_\mathrm{5GHz}/L_\mathrm{2500\text{\AA}})\rt{=2.60\pm0.11}$, where $L_\mathrm{5GHz}$ and $L_\mathrm{2500\text{\AA}}$ are radio and UV monochromatic luminosities at rest-frame 5\,GHz and 2500\,\AA\, 
(see \citealt{Miller2011} for more details). The expected $L_{\mathrm{2keV,RQQ}}$ is derived from the linear relation between X-ray and UV luminosities by \cite{Just2007}. We assume that $L_{\mathrm{2keV,RLQ}}$ represents the observed X-ray luminosity of ID830, while $L_{\mathrm{2keV,RQQ}}$ corresponds exclusively to the \rt{jet-subtracted} X-ray luminosity of ID830. Then, we calculate the X-ray luminosity and $\alphaox$ after removing the jet contamination. \rt{We note that this is a rough approximation to scale down the entire X-ray spectrum, and it neglects the energy dependence of the jet-linked X-ray emission arising from SSC or EC processes.} Table~\ref{tab:alphaOXRQQver} presents the reduction in the X-ray luminosity and $\alphaox$. 

Based on this assumption, we plot the intrinsic \rt{jet-subtracted} X-ray spectrum and $\alphaox$ in Figure~\ref{fig:multiwave_SED} and Figure~\ref{fig:alphaOX}, respectively. The \rt{jet-subtracted} $\alphaox$ shown in Figure~\ref{fig:alphaOX} appears to be consistent with the radio-loud quasars with X-ray excesses described by \cite{Miller2011}. However, considering ID830 to be the super-Eddington quasar, the value of $\alphaox$ is still high compared to those of the PHL 1811 analogs, which are previous candidates for super-Eddington quasars. Similarly, although a slight decrease can be seen, the spectrum shown in Figure~\ref{fig:multiwave_SED} still indicates an excess in X-rays, especially in the soft X-ray band, compared to the blazar sequence with the $46 < \log L_\mathrm{\gamma} < 47$ bin. Therefore, we conclude that the X-ray excess of ID830 cannot be fully explained by the jet-linked components alone. 

\setlength{\topmargin}{3.0cm}
\floattable
\renewcommand{\arraystretch}{1.1}
\begin{deluxetable*}{ccccc}
\thispagestyle{empty}
  \tablecaption{X-ray properties of ID830 after removing the jet contamination.}\label{tab:alphaOXRQQver}
  \tablehead{
        \colhead{$\frac{L_{\mathrm{2keV,RLQ}}}{L_{\mathrm{2keV,RQQ}}}$} & \colhead{$\log L_{\mathrm{2keV}}$} & \colhead{$\log L_{\mathrm{2keV,RQQ}}$} & \colhead{$\alphaox$} & \colhead{$\alpha_\mathrm{OX,RQQ}$} \\
         & \colhead{$(\mathrm{erg\,s^{-1}\,Hz^{-1}})$} & \colhead{$(\mathrm{erg\,s^{-1}\,Hz^{-1}})$} & &
         }
    \startdata
\thispagestyle{empty}
$3.64 \pm 0.12$ & $27.94 \pm 0.14$ & $27.38 \pm 0.14$ & $-1.201 \pm \rt{0.067}$ & $-1.416 \pm \rt{0.067}$ \\
    \enddata
    \thispagestyle{empty}
\tablenotetext{}{
Notes.--- Comparison of the X-ray luminosity at rest-frame 2\,keV and $\alphaox$ before and after removing the contamination from the jet. $\log L_{\mathrm{2\,keV,RQQ}}$ and $\alpha_\mathrm{OX,RQQ}$ represent the \rt{jet-subtracted} X-ray luminosity and $\alphaox$, respectively, assuming no X-ray emission from the jet.
}
\end{deluxetable*}
\setlength{\topmargin}{0in}

\subsubsection{X-ray excess in the slim disk}\label{sec:diskcontamination}

\begin{figure}
\begin{center}
\includegraphics[width=0.48\textwidth]{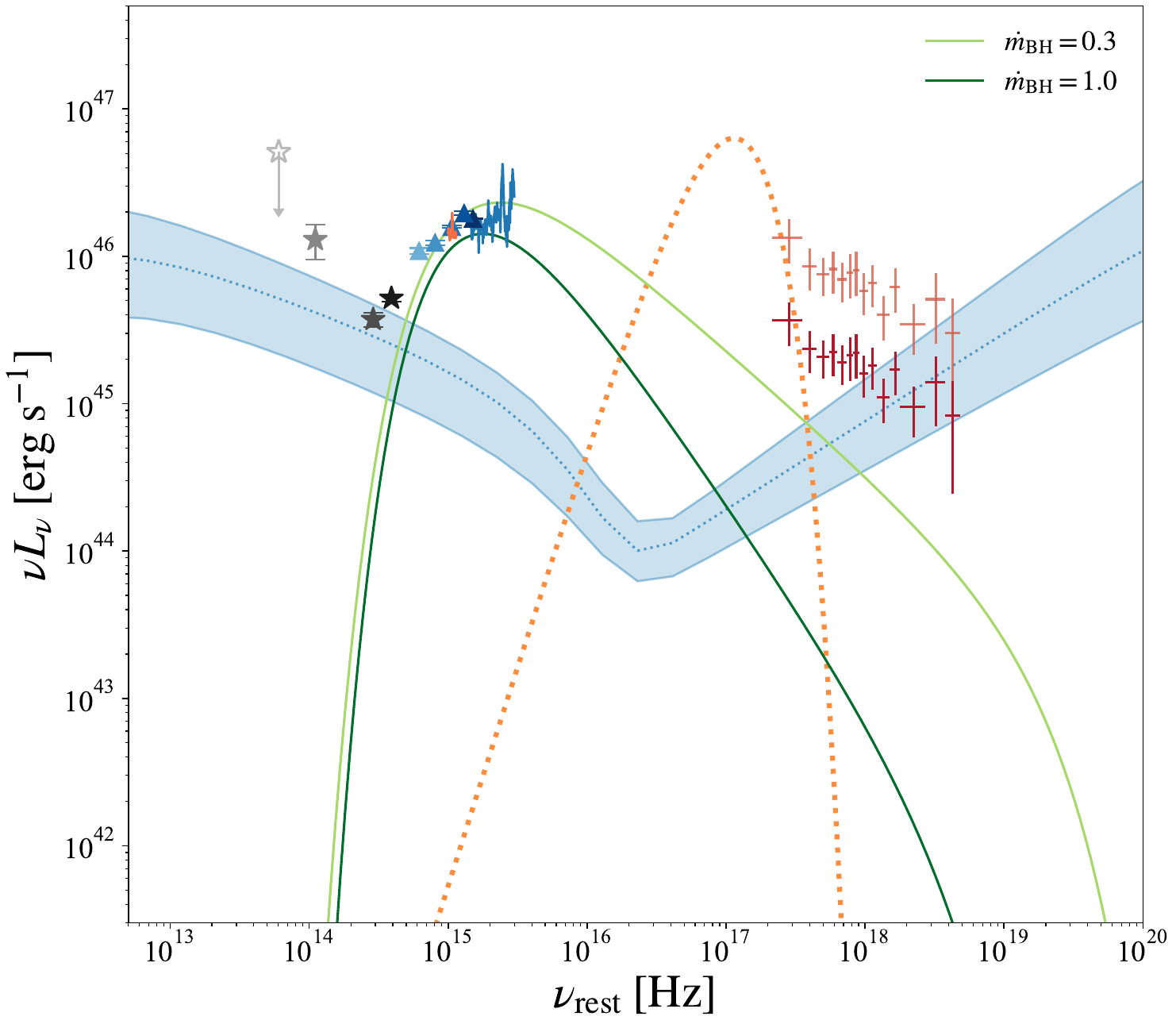}
\caption{
The broad-band SED of ID830 from IR to X-ray with the AGN model SEDs \citep{Inayoshi2024}. 
\rt{The light green and green lines} indicate the SEDs emitted from the warm corona with mass accretion rates of $\dot{m}_\mathrm{BH}=0.3$ and $\dot{m}_\mathrm{BH}=1.0$. They are normalized at their respective peak frequencies. \rt{The blue shaded region represents the blazar sequence in the $46 < \log L_\mathrm{\gamma} < 47$ bin, which is most consistent with the SED of ID830. 
The orange dotted line indicates a blackbody model spectrum of the soft excess component with $E \gtrsim 0.12$ keV, introduced to bridge the gap between the X-ray and UV spectra of ID830.}
}\label{fig:softexcess_SED}
\end{center}
\end{figure}

Another possibility of the X-ray excess is the `soft X-ray excess', whose origins are highly debated \citep[e.g.,][]{don12,wad24,kaw24,ina25}. Some AGNs are known to show an excess in their SEDs in the soft X-ray band. In particular, Narrow-Line Seyfert 1 (NLS1) galaxies typically have the `soft X-ray excess' below $\sim2$\,keV \citep[e.g.,][]{Boller1996}. While the hard X-ray emission generally arises from the inverse Compton scattering in a hot corona, the origin of the `soft X-ray excess' is not fully understood yet. Several components, such as the inverse Compton scattering \citep[e.g.,][]{Magdziarz1998,Gierliński2004,Done2012}, reprocessing and reflection \citep[e.g.,][]{Crummy2006,Fabian2013}, accretion disk itself \citep[e.g.,][]{Jin2013}, and weak jet \citep{Kara2014}, probably contribute to the soft X-ray emission.

Recent JWST has discovered a large number of very compact broad H$\alpha$ emitters with red optical spectra, called little red dots \citep[LRDs; ][]{mat24}. Although they turn out to be very abundant at $z>4$ \citep{aki24,gre24,kok24,koc25}, 
almost all sources lack detectable rest-frame hard X-ray emission at $E>2$~keV \citep{yue24,juo24,ina24}.
Recent studies have shown that such X-ray deficit originates from the colder, warm corona, which produces inverse Compton scattering dominant in soft X-ray bands. 
The warm Comptonization region is low temperature and optically thick, which is different from the hard X-ray emitting region composed of high temperature and optically thin electrons \citep{kaw24,ina25}. 

The soft excess is particularly prominent in NLS1s with high accretion rates. Under super-Eddington accretion state, the outer region of the accretion disk remains sufficiently cool and maintains a geometrically thin structure similar to the standard accretion disk model, whereas the inner region near the central SMBH becomes puffed up, forming a slim disk \citep[e.g.,][]{Abramowicz1988}. \cite{Jin2017a,Jin2017b} presented the structure of the inner accretion flow in a super-Eddington NLS1, and suggested that the region between the inner accretion disk and the compact hard X-ray corona could be a new soft X-ray corona associated with the geometrically thick (`puffed-up') inner disk region in the super-Eddington source. The soft X-ray corona may produce not only warm Comptonization components but also weak reprocessing and reflections partially due to the puffed-up shape (see \citealt{Jin2017a} for more details).

Since NLS1s tend to have smaller black hole masses, the puffed-up region can be located farther from the hot corona, allowing the soft X-ray region to form in between. However, in quasars with $\Mbh \gtrsim 10^8\, \Msun$, the puffed-up region needs to extend sufficiently close to the corona. Therefore, to verify that this soft excess feature observed in NLS1s can be applied to ID830, we overlay the soft excess spectra on the SED of ID830. 
Figure~\ref{fig:softexcess_SED} shows the IR to X-ray SED of ID830 with the blazar sequence \citep{Ghisellini2017}, and the model SEDs of AGN with soft X-ray excess feature due to warm corona at $E \approx 60$~keV for $\dot{m}_\mathrm{BH}=0.3$ and  $E \approx 13$~keV for $\dot{m}_\mathrm{BH}=1.0$ \citep{Inayoshi2024}. The X-ray spectrum for ID830 is located well above the expected model spectra by cooled corona \citep[light green and green curves; ][]{Inayoshi2024}, and thus the origin of our soft X-ray excess cannot be described by the conventional soft X-ray model alone, mainly due to the energy balance between the X-ray and the seed UV photons fixed by our UV spectra.
To reconcile this gap, we find that a hotter warm corona with $E \gtrsim 0.12$ keV (orange dotted curve) is required, suggesting that a soft excess model with higher coronal temperatures is necessary to explain the X-ray spectrum of ID830.

\subsubsection{$\alphaox$ transition with an accretion burst}\label{sec:alphaoxtransition}

\begin{figure}
\begin{center}
\includegraphics[width=0.48\textwidth]{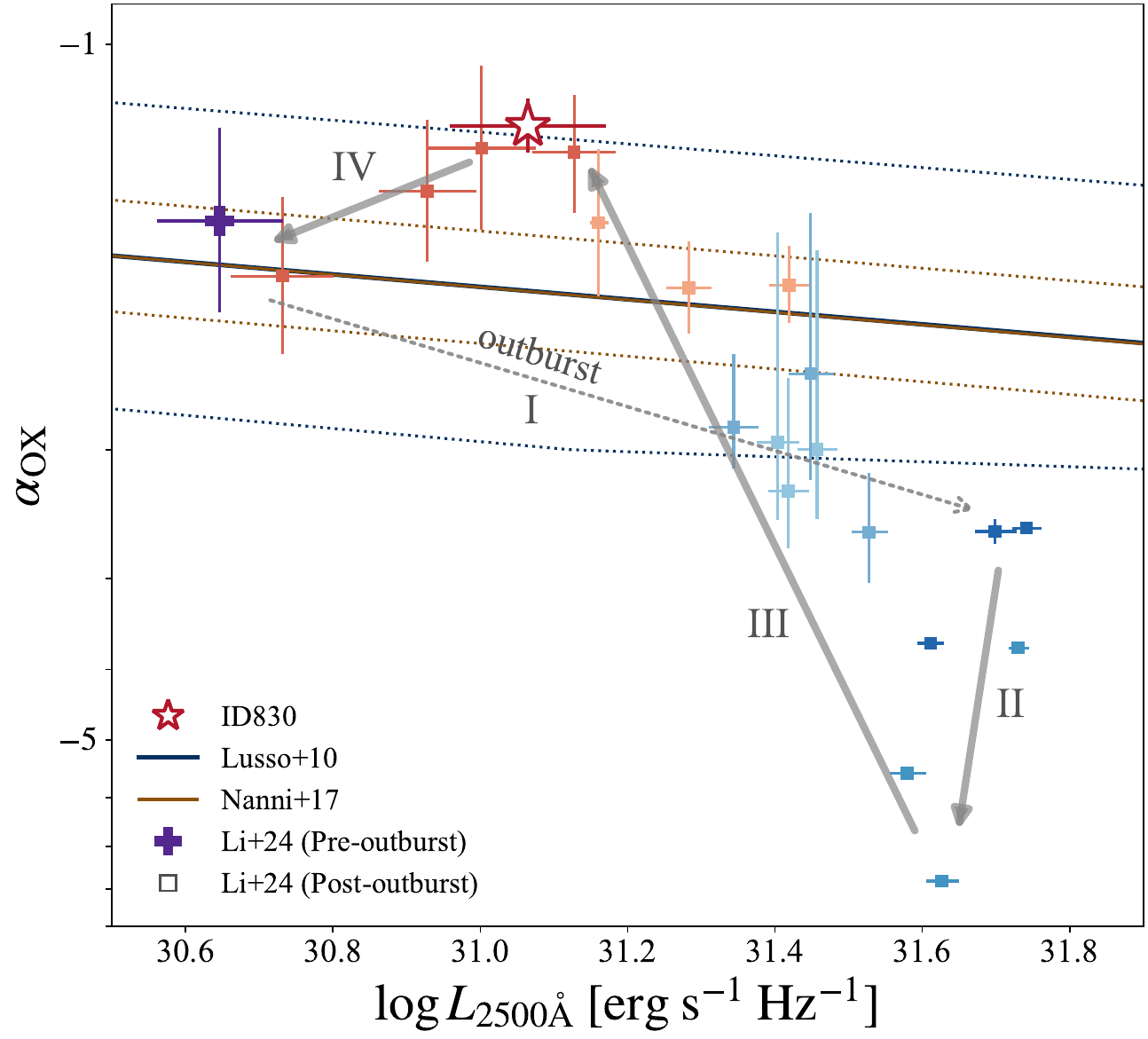}
\caption{
The location of $\alphaox$ for ID830 (red star) and time evolution of $\alphaox$ for changing-look AGN 1ES 1927+654, which experienced a state transition from a super-Eddington phase ($\lambdaedd\approx4$) to a sub-Eddington phase \citep{Li2024}. 
\rt{The best-fit $\alphaox$ relation of \cite{Lusso2010} (blue line) and \cite{Nanni2017} (brown line) are shown and the dotted lines are the $1 \sigma$ upper and lower bounds of the two best-fit relations. 
The UV luminosity of 1ES 1927+654 is shifted along these relations to match ID830. The bold plus marker indicates the pre-outburst value, while square markers with a blue-to-red color gradient represent observed post-outburst data points from a three-year monitoring campaign.} The transient proceeds along the gray arrows.
}\label{fig:alphaOXtransition}
\end{center}
\end{figure}

The further possibility of the X-ray excess of ID830 is the transition of the disk--corona connection driven by a sudden accretion burst.
Recent observations have discovered dramatic flux and spectral variability in AGNs across the entire energy band, especially in the optical/UV and X-ray bands. Such events, known as changing-look AGNs, significantly affect their spectral properties  \citep[e.g.,][]{lam15,Ricci2023}. 
Since they are typically triggered by significant changes in the accretion rate, tracing the state evolution of the accretion disk is effective to understand their structure, and $\alphaox$ serves as a particularly suitable parameter for this purpose. \cite{Li2024} reported time-domain studies of 1ES 1927+654 at $z=0.019422$, which recently underwent a changing-look event, and revealed the transition of its accretion properties after an optical outburst (2017 December 23) by tracking changes in $\alphaox$ over a three-year monitoring campaign. 

\cite{Li2024} described the $\alphaox$ transition in four phases (Phase I to IV), as shown in Figure~\ref{fig:alphaOXtransition}.
Phase I corresponds to the accretion burst. At the beginning of the post-outburst monitoring, 
1ES 1927+654 showed super-Eddington accretion with $\lambdaedd \approx 4$, indicating a slim disk state. 
The most likely scenario suggested by \cite{Trakhtenbrot2019} that induced this rapid increase in mass accretion is a tidal disruption event (TDE; see \citealt{Gezari2021}).
In this extremely super-Eddington state, optically thick outflows strongly enhanced bremsstrahlung cooling due to the increased gas density. Consequently, in Phase II, combined with Compton cooling from the larger number of seed photons, the coronal temperature dropped rapidly, leading to a decrease in the X-ray luminosity. 
In Phase III, although the disk still remained in the super-Eddington state, the mass accretion rate decreased as the TDE material was consumed, causing the slim disk to shrink. As the outflows subsided, the electrons were reheated, leading to a rapid recovery of the coronal temperature and the X-ray luminosity. 
Additionally, the decrease in gas density allowed a larger fraction of energy to be dissipated in the corona. This produces a phase where an excess of X-ray luminosity compared to the decreased UV accretion disk luminosity, reaching $\alphaox \approx -1.0$.
In Phase IV, the mass accretion rate further decreased, causing a transition from the super-Eddington phase to the sub-Eddington phase, i.e., from a puffed-up slim disk to a thin standard disk. As a result, $\alphaox$ returned to its pre-outburst value, reaching a new equilibrium state. The Eddington ratio eventually dropped to $\lambdaedd \approx 0.7$.

Figure~\ref{fig:alphaOXtransition} shows this $\alphaox$ transition of 1ES 1927+654, with the $\alphaox$ value of ID830. We note that the UV luminosity on the horizontal axis is shifted in parallel along the \cite{Lusso2010} relation to match the value of ID830. 
ID830 appears to be precisely in the $\alphaox$ peak phase, corresponding to $\sim 650$ days after the outburst in 1ES 1927+654. 
This suggests that ID830, currently in the super-Eddington state, may be observed in a transitional phase of the disk–corona connection, returning from a burst of mass accretion---potentially triggered by a TDE with a huge gas reservoir \citep[e.g.,][]{ich19c,Kawakatu2025} or other mechanisms---to a sub-Eddington state. Furthermore, the observed $\alphaox$ excess indicates that the corona may have been reheated during this process.

Since 1ES 1927+654 hosts a low-mass black hole with 
$\Mbh=1.9\times10^7\,\Msun$ estimated from the single-epoch method \citep{Trakhtenbrot2019} and is located at low-$z$ ($z=0.019$), its characteristic timescales differ from those of ID830, which hosts the high-mass black hole of $\Mbh=4.40\times10^8\,\Msun$ at high-$z$ ($z=3.4351$). 
The $\alphaox$ transition of 1ES 1927+654 lasts for $t_\mathrm{1ES}\sim 1100$ days \citep{Li2024}. In the case of ID830, the corresponding timescale $t_\mathrm{ID830}$ scales with $\mbh$ and cosmic dilation as
\begin{equation}
    t_\mathrm{ID830} = t_\mathrm{1ES} \times \frac{M_\mathrm{BH,ID830}}{M_\mathrm{BH,1ES}} \times \frac{1+z_\mathrm{ID830}}{1+z_\mathrm{1ES}} \approx 300~\mathrm{yr}.
\end{equation}
Therefore, the transition in ID830 is expected to occur over a much longer timescale. To trace and confirm the $\alphaox$ transition in ID830, future observations targeting larger-scale structures (e.g., the dusty torus, narrow line region; see \citealt{ich19b,pfl22}) will be essential. For example, the dusty torus is known to have a luminosity dependent size of $R_\mathrm{torus} \approx 10 (L_\mathrm{bol}/10^{45} \mathrm{erg\,s^{-1}})^{0.21} \approx 40$~pc for the case of $L_\mathrm{bol,2\text{-}10\,keV}$ \citep{kis11,ich17c}. This corresponds to the light-crossing time $t_\mathrm{lc,torus}\approx 130$~yr, and the observable timescale considering the cosmic dilation would become
$t_\mathrm{delay,torus} \approx 570$~yr. 
Thus, the rest-frame mid-IR emission from the torus would still reflect the previous outburst luminosity, outshining the currently faded UV emission, and future JWST/MIRI observations are essential to confirm this trend.

\subsection{Jet Production in the Super-Eddington Accretion}\label{sec:jetproduction}

Our results show that ID830 hosts accretion disk under the super-Eddington accretion flow with strong radio jet. This indicates that radio jet might has an
important role for AGN feedback even in the high-accretion regime, including super-Eddington phase.

To investigate the energetics of jet in ID830, we calculate the total jet power $P_\mathrm{jet}$ of ID830, using the relation between $P_\mathrm{jet}$ and $L_\mathrm{1.4\,GHz}$ \citep{Cavagnolo2010}:
\begin{equation}
    P_\mathrm{jet} = 1.05 \times 10^{44}~(L_\mathrm{1.4\,GHz} / 10^{24}~\mathrm{W\,Hz^{-1}})^{0.75}\,\mathrm{erg\,s^{-1}}.
\end{equation}

As a result, we obtain $\log (P_\mathrm{jet}/\mathrm{erg}~\mathrm{s}^{-1})=46.47$. The high $P_\mathrm{jet}$ sources with $P_\mathrm{jet} > 10^{45.0}$~erg~s$^{-1}$ are powerful enough to expand the jets to $\gtrsim 10\,$kpc \citep{Nesvadba2017} and promote quenching star formation in host galaxies \citep{McNamara2007}.
As we will discuss later in Section~\ref{sec_sub:LF}, non-negligible fraction
of X-ray luminous quasars might host such radio-jet at $z\sim3$, indicating that AGN feedback would be efficiently working in this early epoch. We note, however, that the $P_\mathrm{jet}$ and $L_\mathrm{1.4\,GHz}$ relation comes with several caveats, including the time evolution of $L_\mathrm{1.4\,GHz}$, large scatter in the slopes, and dependence on additional parameters. Therefore, the discussion above is based on an approximate estimator (see \citealt[Section 6.2.1]{Igo2024}).

\begin{figure*}
\begin{center}
\includegraphics[width=0.90\textwidth]{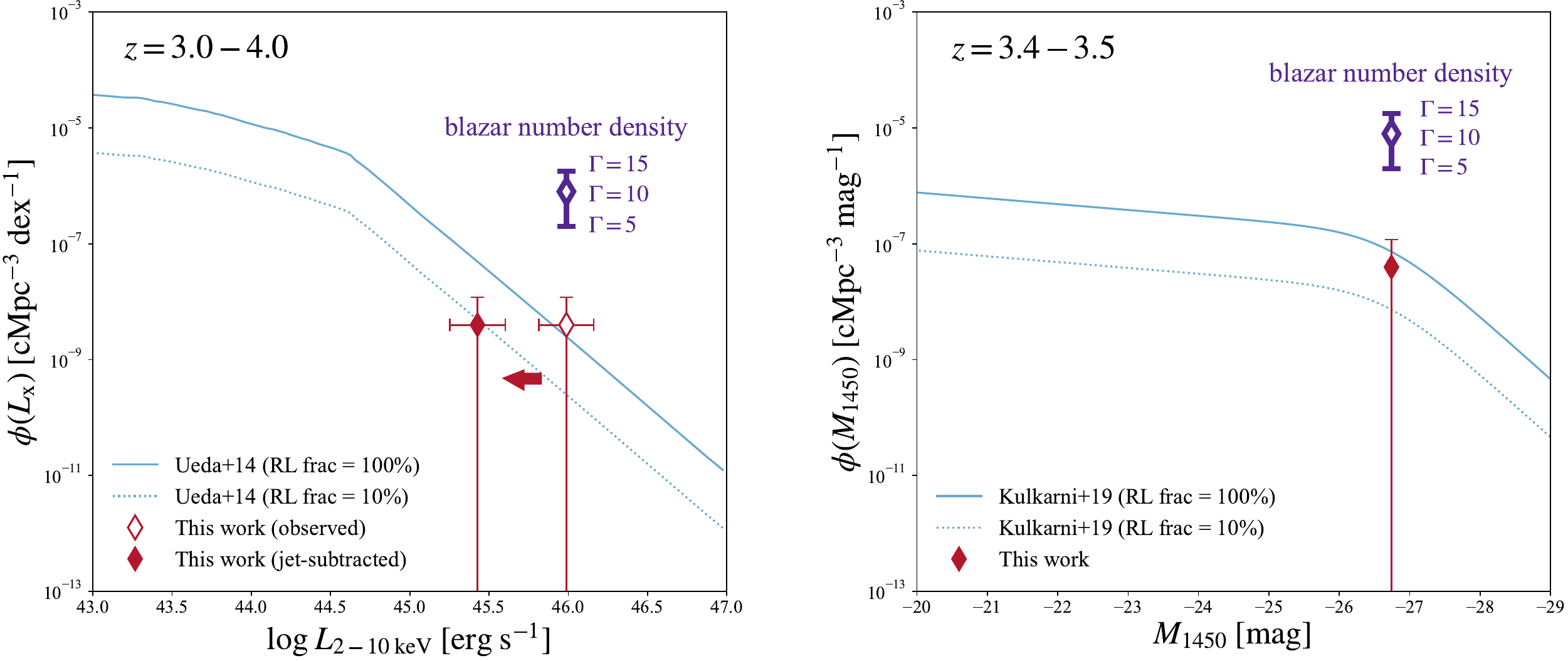}
\caption{
The left panel shows the rest-frame 2--10\,keV X-ray luminosity function \citep{Ueda2014} of AGNs. \rt{The open red diamond represents} the number density plotted at the observed X-ray luminosity $\Lhardx$ of ID830. The filled red diamond represents the number density plotted at the \rt{jet-subtracted} X-ray luminosity, after removing the jet contamination. The right panel shows the rest-frame 1450\,\AA\ magnitude UV luminosity function \citep{Kulkarni2019} of quasars. The filled red diamond represents the number density at the UV magnitude $M_\mathrm{1450\,\text{\AA}}$ of ID830. 
The solid line in each panel indicates the intrinsic luminosity function of the total AGN population, while the dashed line indicates that of the radio AGN population, assuming a radio AGN fraction (RL frac) of 10\%. The purple diamonds are the number densities assuming that ID830 is a blazar. 
\rt{The error bars span the cases for the range for Lorentz factors $\Gamma=5$--15 and do not include the error associated with the number density derived from the observed X-ray luminosity.}
}\label{fig:luminosity_function}
\end{center}
\end{figure*}

There are several theoretical studies on the jet production mechanism in such super-Eddington accretion flow.
For example, \cite{Ohsuga2011} conducted two-dimensional global radiation-magnetohydrodynamic (RMHD) simulations and proposed the global structure of black hole accretion flows and outflows for each accretion mode. 
In the super-Eddington accretion regime, the accretion disk could produce outflows or jets due to the strong radiation force.

Another possibility is that the observed jet in ID830 is produced during a state transition of the accretion disk, from the analogy of the well-studied disk-jet connection in black hole X-ray binaries (BHXRBs; e.g., \citealt{Fender2004, Done2007}). In BHXRBs, three primary accretion states are observationally known based on the accretion rate and X-ray spectral shape \citep[e.g.,][]{vanderKlis1994, Esin1997}: the low/hard state (LS), dominated by a power-law component and typically associated with 
$\lambdaedd \lesssim 0.01$, where steady compact jets are produced; the high/soft state (HS), dominated by thermal emission from the accretion disk, with jets generally suppressed \citep[e.g.,][]{McClintock2006}; and the very high state (VHS), which is more complex and often associated with transient or unstable jet behavior.

\citet{Fender2004} proposed a unified model in which a ``jet line'' separates the jet-producing and jet-suppressed regimes in the hardness–intensity diagram. In particular, transitions across this jet line, often during the soft VHS, can trigger powerful but short-lived radio outbursts.
ID830 shows several characteristics reminiscent of such a transition: a super-Eddington accretion rate ($\lambda_\mathrm{Edd} \gtrsim 1.4$), a steep X-ray photon index ($\Gamma \sim 2.4$), and high jet power ($P_\mathrm{jet} \sim 10^{46.5}$ erg s$^{-1}$). Assuming that the disk–jet coupling observed in BHXRBs also holds qualitatively for AGNs, ID830 may represent a quasar near this jet line, and this assumption is consistent with a proposed state transition phase in Section~\ref{sec:alphaoxtransition}.
Future high-resolution radio observations (e.g., VLBI) could probe the spatial extent of the jet and its structure, potentially constraining the age and duty cycle of the jet activity, and such spatially resolved jet will enable us to estimate more precise jet power.

\subsection{X-ray and UV Luminosity Function of X-ray luminous radio quasars}\label{sec_sub:LF}

We discuss the abundance of radio quasars that possibly contribute to the AGN feedback.
Radio AGNs are known to account for approximately 10\% of the total AGN population \citep{kel94,Ivezić2002}\rt{, although this fraction depends strongly on stellar or black hole mass and redshift \citep[e.g.,][]{Best2005,Rigby2011,hec14,jia16,ich17}.} As mentioned above, we identify at least one X-ray luminous radio-loud quasar in the eFEDS 65\,$\deg^2$ field. In addition, we restricted the sample to sources with spec-$z$. Since the number of spec-$z$ sources drops sharply at fainter magnitudes, they are limited to $\sim 32\%$ of the full sample at the brightness of ID830.
Based on this survey field and the spec-$z$ bias, we estimate the number density of radio-loud quasars.

Figure~\ref{fig:luminosity_function} presents the X-ray luminosity function at $z=3.0$--4.0 \citep{Ueda2014} and the UV luminosity function at $z=3.4$--3.5 \citep{Kulkarni2019}, overlaid with our estimated number densities for radio-loud quasars based on ID830: $n = 3.98 \times 10^{-9}$\,Mpc$^{-3}$ (X-ray-based) and $n = 3.96 \times 10^{-8}$\,Mpc$^{-3}$ (UV-based), each determined within the corresponding survey volume.

In the left panel of Figure~\ref{fig:luminosity_function}, the open red diamond indicates the number density computed using the observed hard X-ray luminosity ($L_{2\text{--}10,\mathrm{keV}}$) of ID830. It lies above the line of radio-loud fraction $=100$\% (cyan solid line), corresponding to $\sim 125$\% of the entire AGN population. \rt{If we correct for jet-linked X-ray contamination as described in Section~\ref{sec:jetcontamination}, the inferred number density (filled red diamond) decreases to $\sim 7.5$\%, in good agreement with the expected radio-loud AGN fraction of $\sim 10$\%. However, since the X-ray luminosity function generally does not consider the jet-linked X-ray enhancement, this result still suggests that the number density of radio-loud AGN may be higher than previously estimated, or that the jet-linked X-ray enhancement in ID830 is exceptionally strong.}

In the right panel of Figure~\ref{fig:luminosity_function}, the UV-based number density also exceeds the nominal 10\% threshold, despite UV luminosity not being significantly boosted by jet emission. The derived value corresponds to $\sim 57$\% of the total AGN population, suggesting that the space density of radio-loud quasars at $z \sim 3$--4 may be underestimated in existing UV selected quasar samples. This discrepancy indicates the importance of wide-area radio quasar surveys to reveal previously undetected radio AGN and quasars, especially at high redshift.

The purple diamonds in Figure~\ref{fig:luminosity_function} represent the inferred number densities under the assumption that ID830 is a blazar. Assuming the Doppler factor $\delta$ is approximately equal to the bulk Lorentz factor $\Gamma$ (i.e., $\delta \sim \Gamma$), the total parent population of blazars is expected to be boosted by a factor of $\sim 2\Gamma^2$, accounting for the relativistic beaming cone of opening angle $\sim 1/\Gamma$. Blazars typically exhibit $\Gamma \sim 10$ \citep[e.g.,][]{Ghisellini2010a, Ghisellini2010c}, so we show the estimated number densities for a plausible range of $\Gamma = 5$--15.

As illustrated, the inferred blazar number densities would significantly exceed the total AGN space density at these redshifts. This discrepancy strongly disfavors the interpretation of ID830 as a blazar. Alternatively, if ID830 were representative of a blazar population, this would imply an overabundance of blazars at high redshift, giving a strong tension in population studies of radio AGN. Previous works have noted that the observed number of high-$z$ blazars appears too large compared to expectations from luminosity functions and jet geometry \citep{Volonteri2011}. 
\rt{If ID830 were a blazer, our findings would provide additional evidence for this discrepancy,} 
further emphasizing the need for improved modeling of radio AGN and their orientation effects at high redshift.

\vspace{3mm}
\section{Conclusion}\label{sec:conclusion}

We have conducted a comprehensive multiwavelength analysis of the radio-loud quasar eFEDS~J084222.9+001000 (ID830) at $z=3.4351$, identified as one of the most X-ray luminous radio quasars discovered in the eROSITA/eFEDS field. By combining eROSITA X-ray spectroscopy, SDSS, and Subaru/MOIRCS rest-frame UV–optical spectra, and extensive radio data from LOFAR, GMRT, FIRST, ASKAP, and VLASS, we have revealed that ID830 represents a rare example of a super-Eddington, radio-loud quasar exhibiting an extreme X-ray excess. Our main findings are:

\begin{enumerate}
\item We identified ID830 as one of the most X-ray luminous radio-loud quasars discovered in the eROSITA/eFEDS field, with $\log L_{2-10\,\mathrm{keV}} = 45.99$~erg~s$^{-1}$ and a steep photon index $\Gamma = 2.43 \pm 0.21$.

\item The optical–NIR spectrum shows moderate reddening ($A_V = 0.39 \pm 0.08$~mag) and a bolometric luminosity estimated from the UV continuum reaches $L_\mathrm{bol} = 7.6\times10^{46}$~erg~s$^{-1}$, implying Eddington ratios $\lambda_{\mathrm{Edd,UV}} \approx 1.4$ and $\lambda_{\mathrm{Edd,X}} \approx 13$ from the X-ray luminosity, confirming super-Eddington accretion.

\item The broadband SED reveals an unusually high $\alpha_{\mathrm{OX}} = -1.20$ (or $-1.42$ after correcting for jet contribution), far above the typical $\alpha_{\mathrm{OX}}$–$L_{2500}$ relation for quasars. This X-ray excess likely arises from both a reheated corona and partial jet-linked inverse Compton scattering. Combined with the analogy of the local changing-look AGN/TDE 1ES 1927+654, our results suggest that ID830 is in a transitional phase in which the corona and jet are simultaneously energized following an accretion burst. ID830 may represent a post-burst super-Eddington quasar bridging the gap between sub-Eddington quasars and the X-ray weak, rapidly accreting ``Little Red Dots'' (LRDs) recently identified with JWST.

\item  The estimated jet kinetic power ($P_{\rm jet} \sim 10^{45}$--$10^{46}$~erg~s$^{-1}$) is comparable to its radiative luminosity, implying that mechanical energy from the jet can efficiently couple to the host interstellar medium. Such kinetic feedback may drive large-scale outflows, heat the circumgalactic gas, and suppress star formation, providing a mechanism linking super-Eddington accretion and the regulation of galaxy growth at $z\sim3$.

\item The UV-based number density of ID830 corresponds to $\sim$57\% of the total AGN population in high luminosity range around $M_{1450}\approx -27$ at $z\sim$3--4, exceeding the expected $\sim$10\% radio-loud fraction, suggesting that radio AGNs like ID830 may be significantly more abundant at high redshift at $z \gtrsim 3$. Given their substantial radio luminosities, such sources could play a non-negligible role in AGN feedback at $z \gtrsim 3$.
\end{enumerate}

\acknowledgments
We thank Kohei Inayoshi for fruitful discussions.
This work is supported by the Japan Society for the Promotion of Science (JSPS) KAKENHI (25K01043; K.~Ichikawa, 23K13154; S.~Yamada).
K.I. also acknowledges support from the JST FOREST Program, Grant Number JPMJFR2466 and
the Inamori Research Grants, which helped make this research possible. Z.I. acknowledges the support by the Excellence Cluster ORIGINS which is
funded by the Deutsche Forschungsgemeinschaft (DFG, German Research
Foundation) under Germany´s Excellence Strategy – EXC-2094 – 390783311.

%

\facilities{VLA, eROSITA, Subaru/MOIRCS, Subaru/HSC, VISTA, WISE, LOFAR, GMRT, ASKAP, \textit{Swift}/BAT, \textit{Fermi}/LAT}

\vspace{8mm}
\bibliographystyle{aasjournal}
\bibliography{eFEDSradio}



\end{document}